\RequirePackage{lineno}
\documentclass[prc,reprint,superscriptaddress]{revtex4-1}
\pdfoutput=1
\usepackage{graphicx}
\usepackage{amssymb}
\usepackage{amsmath}
\usepackage{hyperref}
\usepackage{color}
\usepackage{bm}
\usepackage{color}
\usepackage{lineno}
\usepackage{subfig}

\begin{document}

\title{Monte Carlo Simulations of Trapped Ultracold Neutrons in the UCN\texorpdfstring{\(\tau\)}{tau} Experiment}
\author{Nathan Callahan}
\author{Chen-Yu Liu}
\author{Francisco Gonzalez}
\author{Evan Adamek}
\affiliation{Center for Exploration of Energy and Matter; Physics Department, Indiana University, Bloomington, IN 47408, USA}
\author{James David Bowman}
\author{Leah Broussard}
\affiliation{Oak Ridge National Laboratory, Oak Ridge, TN 37831, USA}
\author{S.M. Clayton}
\author{S. Currie}
\affiliation{Los Alamos National Laboratory, Los Alamos, NM 87545, USA}
\author{C. Cude-Woods}
\affiliation{Department of Physics, North Carolina State University, Raleigh, NC 27695, USA}
\affiliation{Los Alamos National Laboratory, Los Alamos, NM 87545, USA}
\author{E.B. Dees}
\affiliation{Department of Physics, North Carolina State University, Raleigh, NC 27695, USA}
\author{X. Ding}
\affiliation{Department of Physics, Virginia Polytechnic Institute and State University, Blacksburg, VA 24061, USA}
\author{E.M. Egnel}
\affiliation{West Point Military Academy, West Point, NY 10996, USA}
\author{D. Fellers}
\affiliation{Los Alamos National Laboratory, Los Alamos, NM 87545, USA}
\author{W. Fox}
\affiliation{Center for Exploration of Energy and Matter; Physics Department, Indiana University, Bloomington, IN 47408, USA}
\author{P. Geltenbort}
\affiliation{Institut Laue-Langevin, 38000 Grenoble, France}
\author{K.P. Hickerson}
\affiliation{University of California, Los Angeles, CA 90095, USA}
\author{M.A. Hoffbauer}
\affiliation{Los Alamos National Laboratory, Los Alamos, NM 87545, USA}
\author{A.T. Holley}
\affiliation{Department of Physics, Tennessee Technoligical University, Cookeville, TN 38505, USA}
\author{A. Komives}
\affiliation{Department of Physics and Astronomy, DePauw University, Greencastle, IN 46135, USA}
\author{S.W.T. MacDonald}
\author{M. Makela}
\author{C.L. Morris}
\author{J.D. Ortiz}
\author{R.W. Pattie, Jr}
\author{J. Ramsey}
\affiliation{Los Alamos National Laboratory, Los Alamos, NM 87545, USA}
\author{D.J. Salvat}
\affiliation{Department of Physics, University of Washington, Seattle, WA 98195-1560, USA}
\author{A. Saunders}
\author{S.J. Seestrom}
\affiliation{Los Alamos National Laboratory, Los Alamos, NM 87545, USA}
\author{E. I. Sharapov}
\affiliation{Joint Institute for Nuclear Research, Dubna, Moscow 141980, Russia}
\author{S.K.L. Sjue}
\author{Z. Tang}
\affiliation{Los Alamos National Laboratory, Los Alamos, NM 87545, USA}
\author{J. Vanderwerp}
\affiliation{Center for Exploration of Energy and Matter; Physics Department, Indiana University, Bloomington, IN 47408, USA}
\author{B. Vogelaar}
\affiliation{Department of Physics, Virginia Polytechnic Institute and State University, Blacksburg, VA 24061, USA}
\author{P.L. Walstrom}
\author{Z. Wang}
\author{H. Weaver}
\author{W. Wei}
\affiliation{Los Alamos National Laboratory, Los Alamos, NM 87545, USA}
\author{J. Wexler}
\author{A.R. Young}
\author{B.A. Zeck}
\affiliation{Department of Physics, North Carolina State University, Raleigh, NC 27695, USA}

\date{\today}

\begin{abstract}
 	In the UCN$\tau$ experiment, ultracold neutrons (UCN) are confined by magnetic fields and the Earth's gravitational field. 
	Field-trapping mitigates the problem of UCN loss on material surfaces, which caused the largest correction in prior neutron experiments using material bottles.  
	However, the neutron dynamics in field traps differ qualitatively from those in material bottles.
	In the latter case, neutrons bounce off material surfaces with significant diffusivity and the population quickly reaches a static spatial distribution with a density gradient induced by the gravitational potential. 
In contrast, the field-confined UCN---whose dynamics can be described by Hamiltonian mechanics---do not exhibit the stochastic behaviors typical of an ideal gas model as observed in material bottles.
	In this report, we will describe our efforts to simulate UCN trapping in the UCN$\tau$ magneto-gravitational trap. We compare the simulation output to the experimental results to determine the parameters of the neutron detector and the input neutron distribution. The tuned model is then used to understand the phase space evolution of neutrons observed in the UCN$\tau$ experiment. We will discuss the implications of chaotic dynamics on controlling the systematic effects, such as spectral cleaning and microphonic heating, for a successful UCN lifetime experiment to reach a 0.01\% level of precision.

\end{abstract}

\maketitle

\section{Overview}
\begin{figure*}
	\includegraphics[width=0.75\textwidth]{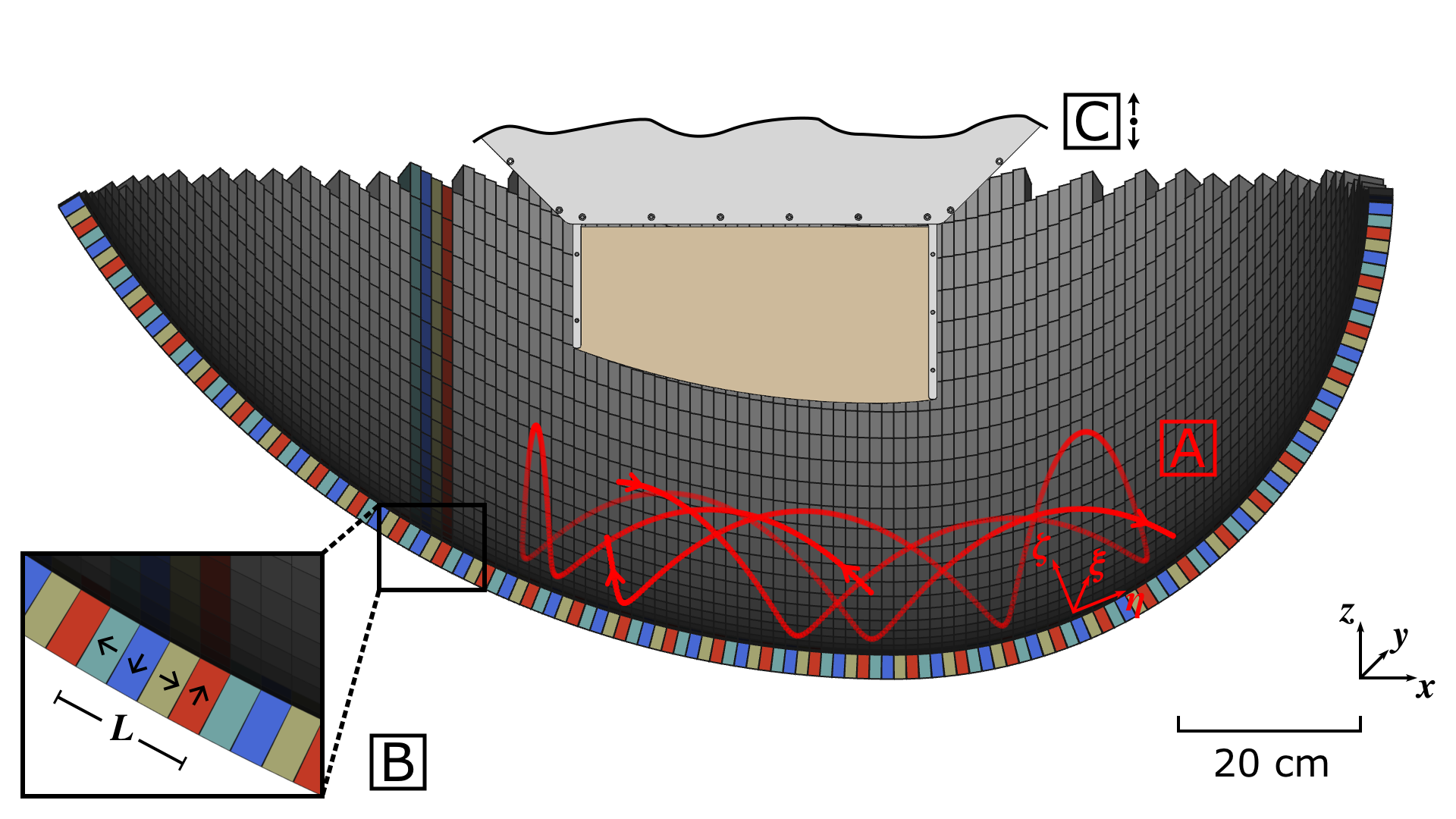}
	\caption{A cross-sectional view of the UCN\(\tau\) magneto-gravitational trap. $\boxed{\textsf{A}}$ illustrates an example UCN trajectory inside the trap. The lab coordinate system $x$-$y$-$z$ and the local coordinate system $\xi$-$\eta$-$\zeta$ are shown. $\boxed{\textsf{B}}$ shows a cross-section of rows of magnets with identical magnetization highlighted by the colored strips. Arrows indicate the magnetization direction for each row of magnets.  $\boxed{\textsf{C}}$ shows the \textit{in-situ} detector, referred as the dagger detector, which can be  to moved vertically in and out of the trap volume. The beige-colored region is the active area of the neutron detector, coated with \(^{10}\)B, and the gray is the detector housing and support structure. (color online)}
	\label{fig:trajintrap}
\end{figure*}
The UCN$\tau$ experiment is designed to measure the lifetime of neutrons by
counting the number of surviving ultracold neutrons (UCN) in a trap~\cite{salvat, morrisrsi, robby}.
The experiment is motivated by the discrepancy in the values of neutron lifetime measured by the methods of beam and bottle; see reviews~\cite{Nico2009, serebrov2011, Wietfeldt2011, Dubbers2011, Greene2016} for details.
The UCN$\tau$ experiment uses a magneto-gravitational trap to confine UCN in an open-top magnetic bowl. A permanent magnet array in a Halbach configuration produces a steep magnetic gradient near the array surface.
About 5,300 discrete permanent magnets, each with a pole strength of 1~T, were used in forming the magnetic array. The array covers the inside of a large concave bowl as shown in Fig.~\ref{fig:trajintrap}. 
The resulting magnetic field decays exponentially away from the inner surface of the bowl. 
Polarized neutrons (in the low-field-seeking state) bounce off the magnetic field of the bowl, rise up vertically, and are pulled back down into the bowl by gravity.
As such, gravity forms the lid of this open-top trap.
In this magneto-gravitational (MG) trap, UCN with kinetic energy below 50~neV are completely repelled by the magnetic field; they rise to a maximum height of 0.5~m in the Earth's gravitational field. They are stored in the trap without experiencing material interactions and the associated losses. Additionally, a holding field everywhere perpendicular to the trapping field is provided to ensure that there are no regions of vanishing magnetic field inside the trapping volume and depolarization losses are negligible~\cite{walstrom,steyerl2017}. An {\it in-situ} neutron detector, using $^{10}$B coated ZnS scintillating sheets~\cite{zwang} (shown in Fig.~\ref{fig:trajintrap}C), can be moved into the trap to count the surviving neutrons.

In this paper, we present Monte-Carlo simulations to study the dynamics of these field-trapped UCN.  Similar work can be found in Ref.~\cite{serebrov2008, serebrov2018} with material bottles and in Ref.~\cite{coakley2005, bowman2005, walstrom, berman, Bowman2014, Liu2014} for field traps.
This work attempts to understand 
the behavior of the UCN population in the UCN$\tau$ MG trap---where each neutron reflection is deterministic and non-diffusive---and its effects on the neutron counting efficiency using the {\it in-situ} detector. 
The physics model of the surface interaction used in material bottles (with the Fermi-potential, the diffusivity and the loss per bounce as tunable parameters~\cite{serebrov2008}) is now replaced by a field interaction model~\cite{walstrom}.
Since the field is known (or can be measured to a high degree of precision), there is little uncertainty in the micro-physics of neutron scattering.  
On the other hand, the reduced number in the degrees of freedom makes it challenging to construct a Monte-Carlo simulation that reproduces the experimentally measured data. 
Nevertheless, the simulations elucidate a nonuniform geometrical acceptance of the overthreshold neutrons by the {\it in-situ} detector. 
Overthreshold neutrons are neutrons with kinetic energies larger than the trapping potential, but could reside in certain quasi-stable orbits and remain in the trap during the finite measurement time.
The nonuniform geometrical acceptance of these neutrons may imply that the procedure we used in Ref.~\cite{morrisrsi, robby}---based on the counts measured at the cleaning height to constrain the systematic effects of spectral cleaning and heating---requires some refinement.
The size of these systematic effects, on the other hand, is reasonably well constrained by the Monte-Carlo simulations.
All of the simulations indicate that the cleaning procedures in place put stringent bounds on possible systematic errors due to untrapped neutrons and heating.  Reproducing the arrival time data for detected neutrons in detail is more challenging.  This requires fine-tuning a relatively large set of correlated parameters to reproduce the measured spectra. Although the level of success is encouraging, it is clear that further development of both measurements and the simulations presented here are required to produce convincing agreement between simulations and measurement, especially when treating the subtle effects connected to phase space evolution. 

The paper contains two parts. The first part discusses the physics models and the optimization of input parameters by comparing to experimentally acquired data on the neutron arrival time.
Details of the trapping potential, the numerical integration, and the neutron detection are presented in Sec.~\ref{sec:Simulation}; the data analysis in Sec.~\ref{sec:Post-simulation Analysis and Model Optimization}; the optimization of model parameters in Sec.~\ref{subsec:chisqmin}. 
The second part discusses the neutron dynamics. 
The chaotic motions and their implications for spectral cleaning are discussed in Sec.~\ref{sec:chaos}. The effects of neutron heating due to microphonic vibration and the estimates of the systematic shift in the neutron lifetime are presented in Sec~\ref{sec:p1corr}.

\section{Simulation} \label{sec:Simulation}
Each simulation tracks about $10^5$ to $10^6$ neutrons in the trap, by numerically integrating the equations of motion. The field potential, following previous work ~\cite{walstrom, berman}, is described by a Halbach array field expansion using a local coordinate system on a curved surface. Details of the numerical integration are presented in Appendix~\ref{sec:appendix}, along with data testing the numerical integrations, including the degree of energy conservation, the step size selection, the expansion truncation, and the numerical convergence. We will start by discussing how we model the neutrons in the MG trap (see Fig.~\ref{fig:trajintrap}) used in the UCN$\tau$ experiment.

\subsection{UCN event generation, tracking, and detection} \label{sec:ucngen}
Because the neutrons enter the trap through a removable segment of the halbach array (the trapdoor) located in the bottom of the trap, 
UCN were generated randomly on a 15~cm\(\times\)15~cm plane (the size of the trapdoor opening) placed at the height of the zero potential point at 3.5~cm above the bottom of the trap. 
The difference between the energy and the potential at birth is the kinetic energy, which sets the  initial velocity, $\sqrt{2(E-V)/m_n}$, points along an azimuthal angle $\phi_0$ and a polar angle \(\theta_0\) relative to $z$.  Both angles are generated at random.
A linear energy distribution is used for studies where the spectrum of the trap is tuned in the post-simulation analysis; otherwise a parameterized energy distribution is used.
UCN were generated up to an energy of 46.1~neV, which is sufficiently larger than the energy needed to reach the nominal height of the spectral cleaner (38~cm).
After the initialization, each neutron is tracked for a dwelling time $t_d$ representing the filling process, followed by a period of time, $t_c$, for spectral cleaning.
The time constant to fill an empty UCN$\tau$ trap, measured after the trap door opens, is about 70~s.
During this time, a constant flow of neutrons enter the trap; as the trapdoor stays open, many neutrons already entered the trap exit just as easily. Filling the trap from below with an open trapdoor may condition the phase space distribution inside the trap and affect the fraction of semi-periodic neutrons.
However, it is difficult to model this filling stage, as the magnetic field with an open door is significantly more complicated and has not yet been modeled.
The current model only implements the magnetic field of a closed trap. In order to simulate the filling procedure, in which each neutron would enter the trap at a different time, we set a dwelling time for each UCN as exponentially distributed with a time constant of 70~s and truncated at 150~s. 
After the dwelling time,  $t_d$, each UCN is subject to 50~s or 200~s of spectral cleaning to emulate the dataset of interest.
After the cleaning period, the cleaner is immediately moved to a height of 43~cm and remains a perfect absorber in the trap.

After the initial combined dwelling and cleaning time, $t_d+t_c$,  
the UCN are tracked
throughout the duration of the preset storage time
between 20~s and 1400~s (that matches our experimental procedure).
Neutron loss through \(\beta\) decay is generally included during the tracking. In some cases, it was added later in the post-simulation analysis.
Finally, the detector is simulated by inserting at a speed of 3.8~cm\,s\(^{-1}\) and stopping at the programmed heights of the simulated run.

\begin{figure}
	\includegraphics[width=0.45\textwidth]{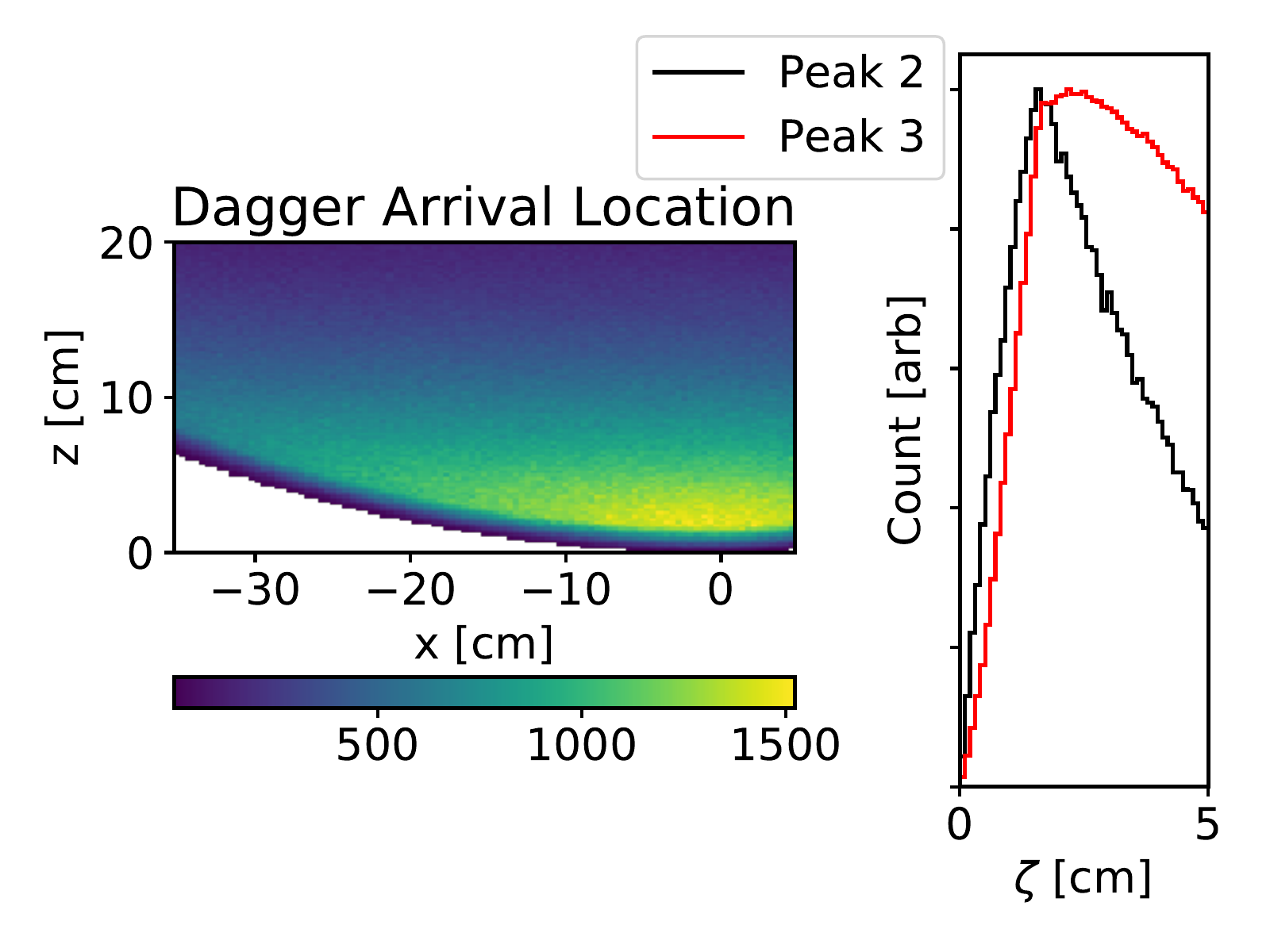}
	\caption{Simulation of neutron hits on the dagger detector. The left contour plot shows the distribution of arrival locations on the surface of the dagger detector summed over all detector positions. The right plot shows the probability of a UCN hit as a function of $\zeta$, the distance from the bottom edge of the dagger detector, in a 3-step measurement. Note that the neutrons detected in the lowest position (peak 3) are distributed higher than that in peak 2.}
	\label{fig:DaggerHit}
\end{figure}

The dagger detector (see Fig.~\ref{fig:trajintrap}C and Fig. ~\ref{fig:DaggerHit} left) is on the mid-plane \(y=0\), with a cross-section defined by
\begin{eqnarray}
-20\text{ cm}&<&(x-x_{off})< 20\text{ cm,  and}  \nonumber \\ 
f(x)&<&z<h(t)+20\text{ cm},
\label{eq:DaggerProfile}
\end{eqnarray}
where $x_{off}=15.24$~cm is the position offset of the dagger detector from the origin of the lab coordinate system, and \(f(x)\) is the curved profile of the bottom of the dagger detector that matches the curved inner surface of the Halbach array. \(f(x)\) is set at \(\zeta'=0\) where \(\zeta'\) is the local bowl normal offset by the detector height. The lowest edge of the dagger detector, $h(t)$, defines the height of the detector.
In addition, the housing above the neutron-active part of the dagger detector (which contains the fibers, the photomultipliers and cooling loops) is simulated as a box made out of aluminum. It has two parts:
The lower housing is a trapezoid whose lower width is 40~cm, upper width is 69.215~cm, and height is 14.478~cm. The upper housing is a rectangle of width 69.215~cm and height 12.192~cm.
Every time a UCN trajectory crosses the \(y=0\) plane (indicated by a sign change in the $y$-coordinate), the approximate crossing point \((x,z)\) is extrapolated to determine whether it falls within the detector cross-section.
The simulated position distribution of the neutron hit on the dagger detector (for all neutron events recorded in a 3-step measurement) is illustrated in Fig.~\ref{fig:DaggerHit}. 
This hit profile matches the shape given in Eq.~\ref{eq:DaggerProfile}.
Note that the majority of the neutron hits are registered within the 5~cm of the bottom edge.
In a 3-step measurement, the detector is first inserted to measure the remaining neutrons at the cleaning height, followed by lowering in stages to detect UCN at subsequently lower energies. The arrival time profile of one of these steps is referred to as a peak; peak 1 corresponds to the check for uncleaned neutrons and peaks 2 and 3 to detection of trapped UCN.
As shown in Fig.~\ref{fig:DaggerHit} right, the neutrons detected in peak 3 are distributed at a larger $\zeta$ relative to those in peak 2. In peak 3, the bottom of the detector is only 1~cm from the Halbach array and UCN are repelled by the strong magnetic field to reach higher positions.

For each detector interaction, we either simulate absorption during tracking or during post-processing. In the former case, we record the first successful absorption time, energy, and location. In the latter case, we record the first 50 interaction times and perpendicular energies. The latter case is used when tuning the detector model.
If a UCN is not absorbed at the detector surface, the neutron is reflected to a different angle to simulate a non-absorbing surface interaction.
Upon reflection, the outgoing angle follows a Lambertian distribution, i.e., the azimuthal angle is uniformly distributed in $2\pi$, and the polar angle follows a probability of $\sin^{-1}(\sqrt{U})$, where $U$ is a uniform random number between [0, 1].
Similarly, a UCN is scattered diffusively using a Lambertian model if it intersects the aluminum detector housing.

\section{Post-simulation Analysis and Model Optimization} \label{sec:Post-simulation Analysis and Model Optimization}
During tuning of the Monte Carlo, the record of the simulated detector interactions is processed, using the surface model of interest, to generate arrival time histograms in the following fashion.
Each UCN, $n_i$, is given a weight based on the spectrum model \(w\).
Next, for each of the 50 recorded detector interactions $j$ of this neutron, a surface model is invoked to determine the probability of neutron absorption \(\mu(E_{\perp\:i,j})\) in the $^{10}$B surface layer; a uniform random number \(u_{i,j}\) is generated for each detector interaction for the given UCN. 
The time of the detector hit event, for the given UCN, is set to be the first detector interaction where \(u_{i,j} < \mu(E_{\perp\:i,j})\). 
Once the neutron is absorbed in the detector, the record of subsequent detector interactions is ignored. In this way, many detector models can be tested using the same simulation output. 

\subsection{Post-simulation Spectral Weighting} 
Even though the energy spectrum of the UCN output from the LANL source has been measured, the initial UCN energy spectrum in the UCN$\tau$ trap is not well known,
as the geometry and the material choice of the UCN guides together with the vertical position of the trap modify (or filter) the spectrum of neutrons entering the trap.
Simulated neutrons are genered by sampling from a superthermal energy spectrum~\cite{golub} and lambertian angular distribution and are then weighted by
$w(E,\theta_0)\propto E^{x'-1}\text{cos}^{y'}(\theta_0)$
to present a parametrized UCN distribution of 
\begin{equation}
\rho(E,\theta_0)\propto \Theta(E-E_\text{cut})E^{x'}\text{sin}(\theta_0)\text{cos}^{1+y'}(\theta_0),
\label{eq:spectrum}
\end{equation}
where \(x'\) gives the scaling of density with UCN energy, \(y'\) models the forward directedness of the UCN flux, \(\theta_0\) is the initial polar angle of UCN, and \(E_\text{cut}\) is the low energy cut-off, due to the fact that low-energy neutrons are not reflected by the trap door in the open position. 
Inside a superthermal source, UCN are produced with spectral density $x'=1$.  
For neutrons entering the UCN$\tau$ trap (after several meters of neutron guides), this power-law scaling is adjustable to allow for spectral distortion due to the energy-filtering during the neutron transport.
The UCN fill the UCN$\tau$ trap with the trapdoor open. 
Since the trapdoor is positioned at the bottom of the trap,
low-energy neutrons, which populate the region in the immediate vicinity of the trapdoor, readily exit the trap. 
As such, a low-energy cutoff was assumed in the UCN spectrum. 

\subsection{Detector Surface Model}
The rate of the detected neutron events depends on the details of the surface interaction on the detector. We explore a variety of detector models in the post-simulation analysis.
Ideally, the dagger detector has a 
flat UCN surface composed of three parts: a top layer of B\(_2\)O\(_3\), followed by a layer of pure \(^{10}\)B, and an underlayer of bulk ZnS.
The top oxide layer allows variations of the effective Fermi-potential.  
The thickness of the \(^{10}\)B layer controls the probability of absorption for each UCN interaction.
The ZnS crystals are around 10~\(\mu\)m in diameter~\cite{zwang}, which is larger than the UCN wavelength ($\sim$ 500~\AA).
Therefore, the use of the simple flat layer model is justified to estimate the probability of reflection and absorption.
The polycrystalline nature of ZnS, however, will increase diffuse reflection that randomizes the direction of the reflected neutrons.

\begin{figure}
	\includegraphics[width=0.45\textwidth]{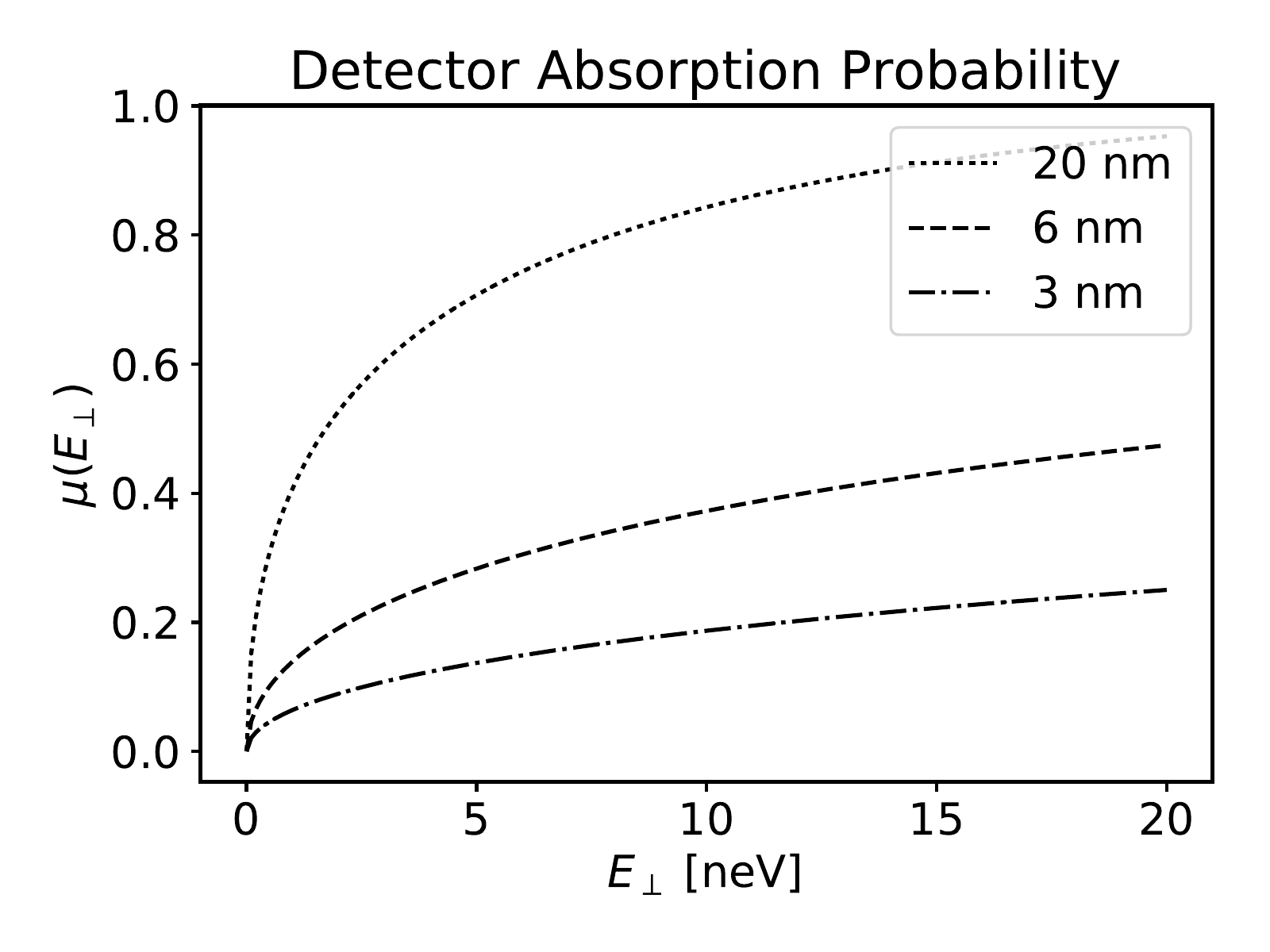}
	\caption{The neutron absorption probability as a function of the neutron energy (perpendicular to the detector plane) for a $^{10}$B layer of 3, 6 and 20~nm.}
	\label{fig:mu}
\end{figure}

To estimate the absorption probability for a multilayer system, we follow the treatment given in Appendix 4 of \cite{golub}. 
Adopting 1-$d$ quantum mechanical step potentials, the reflection coefficient \(R\) on the first incident boundary can be estimated by
\begin{equation}
R=\frac{-\bar{M}_{21}}{\bar{M}_{22}},
\label{eq:R}
\end{equation}
where \(\bar{M}=\bar{M^N}\ldots\bar{M^2}\times\bar{M^1}\) is the product of matrices which match the boundary conditions on each of the \(N\)-th layer boundaries with \(\bar{M}^{N}_{ij}\) as the coefficients for the transmitted and reflected wave (\(j\)) or the wavefunction and its derivative (\(i\)).
For the neutrons remaining inside the layers, they are mostly likely absorbed and subsequently generate a scintillation signal inside ZnS; this gives the absorption probability of
\begin{equation}
\mu(E_\perp)=1-\lvert R \rvert^2.
\label{eq:mu}
\end{equation}
The absorption probability for a nominal detector surface is plotted in Fig.~\ref{fig:mu}.

\begin{figure}
	\includegraphics[width=0.45\textwidth]{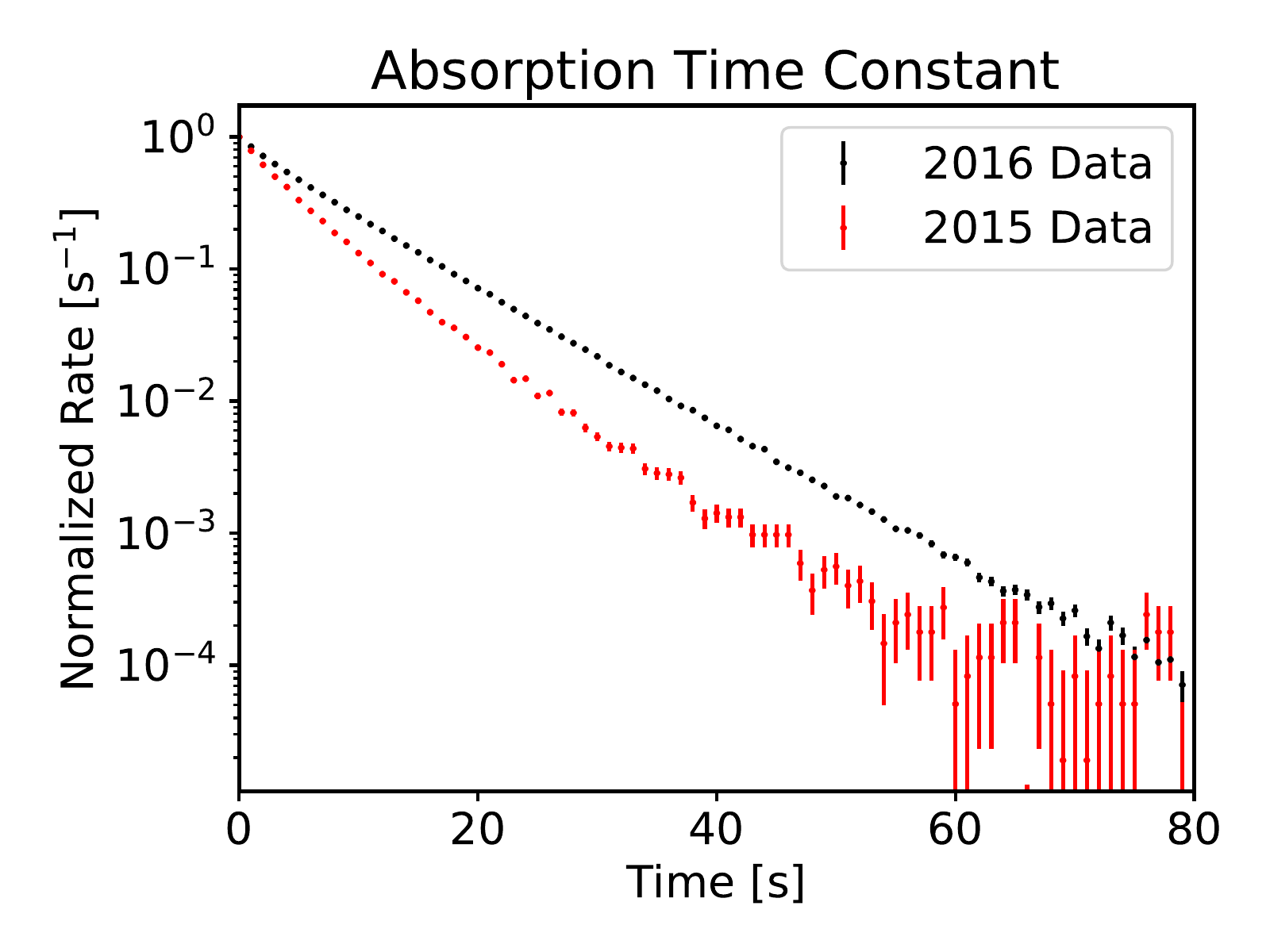}
	\caption{Neutron counted by the {\it in-situ} detector as a function of time after the detector insertion. Note that the same detector shows different time constants for two subsequent years of operation.}
	\label{fig:AbsProbByYear}
\end{figure}

\begin{figure}
	\includegraphics[width=0.45\textwidth]{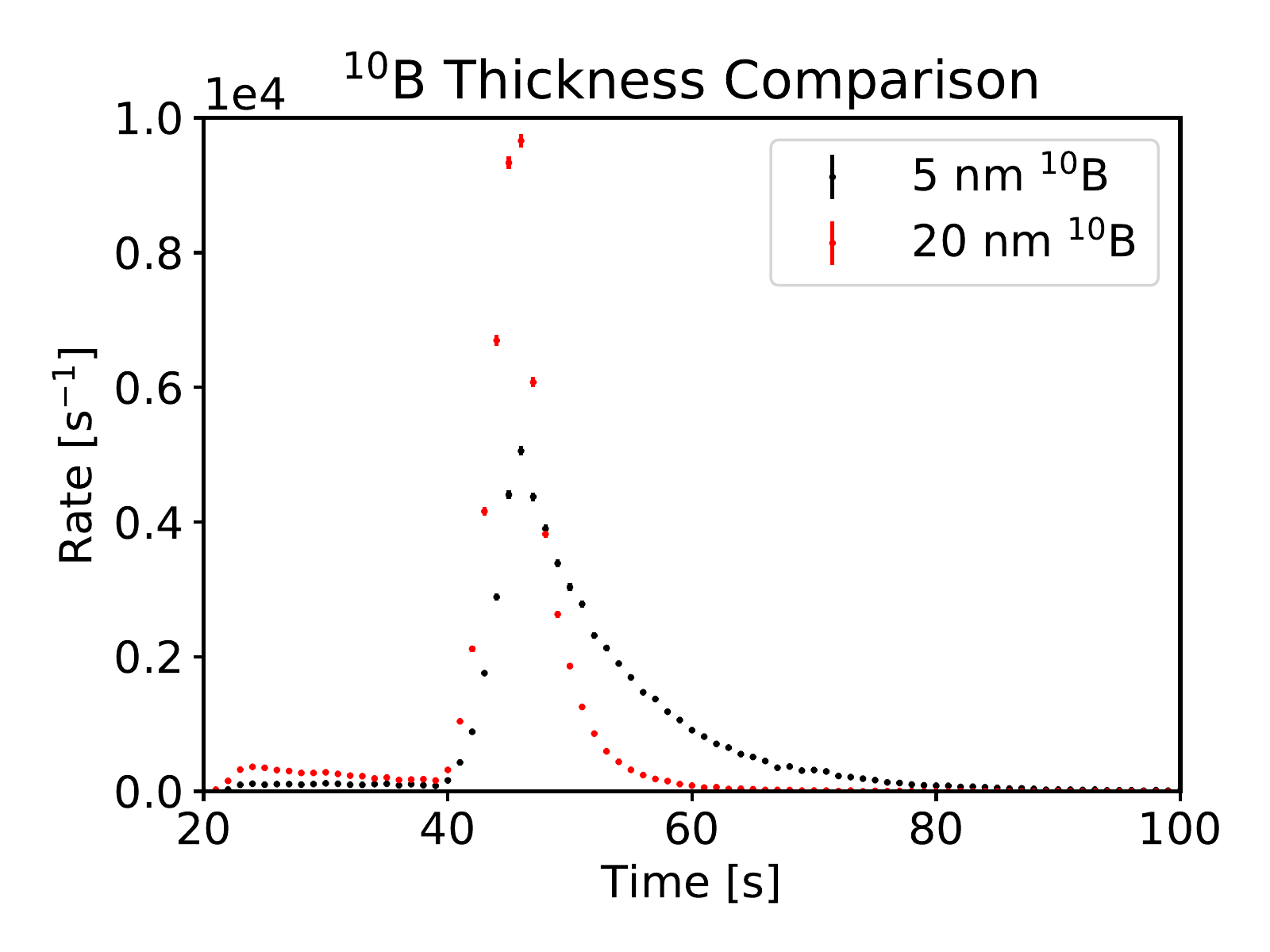}
	\caption{Detector response (for a 3-step measurement) with two different thicknesses of the boron layer. The one with a thicker boron layer counts the neutrons in the MG trap faster. }
	\label{fig:BoronThickness}
\end{figure}

Data from consecutive run cycles exhibited an increasingly longer absorption time,  shown in Fig.~\ref{fig:AbsProbByYear}.  Three effects could explain the observed increase: an oxide or contamination layer building up over time, a change in the thickness of the boron (see Fig.~\ref{fig:BoronThickness}), or damage to the detector from occasional mechanical malfunctions over the course of the experiment. 
Mechanical damage was caused by dropping the detector onto the halbach array, resulting in visible damage on the $^{10}$B-coated surface near the bottom of the dagger detector.

We model the effect of this detector damage
by modifying Eq.~\ref{eq:mu} with an independent probability of interacting with a bare patch instead of \(^{10}\)B. 
If we assume that the damage is limited to the bottom band with a thickness of \(\zeta_\text{cut}\), then
the modified probability of absorption is modeled as
\begin{equation}
P(E_\perp,\zeta')=\begin{cases}
\frac{\zeta'}{\zeta_\text{cut}}\times\mu(E_\perp) & 0 \leq \zeta' < \zeta_\text{cut} \\
\mu(E_\perp) & \zeta' \geq \zeta_\text{cut}, \\
\end{cases}
\label{eq:absprob}
\end{equation}
where \(\zeta'\) is the distance from the bottom edge of the detector. Incidentally, UCN absorbed during the last step have a larger \(\zeta\) on average than in higher steps due to the large field below the detector, and thus a larger probability of absorption, resulting in a faster draining time. 

\subsection{Parameter Tuning via the \texorpdfstring{\(\chi^2\)}{chi square} Minimization} \label{subsec:chisqmin}
\begin{figure*}
    \includegraphics[width=0.65\textwidth]{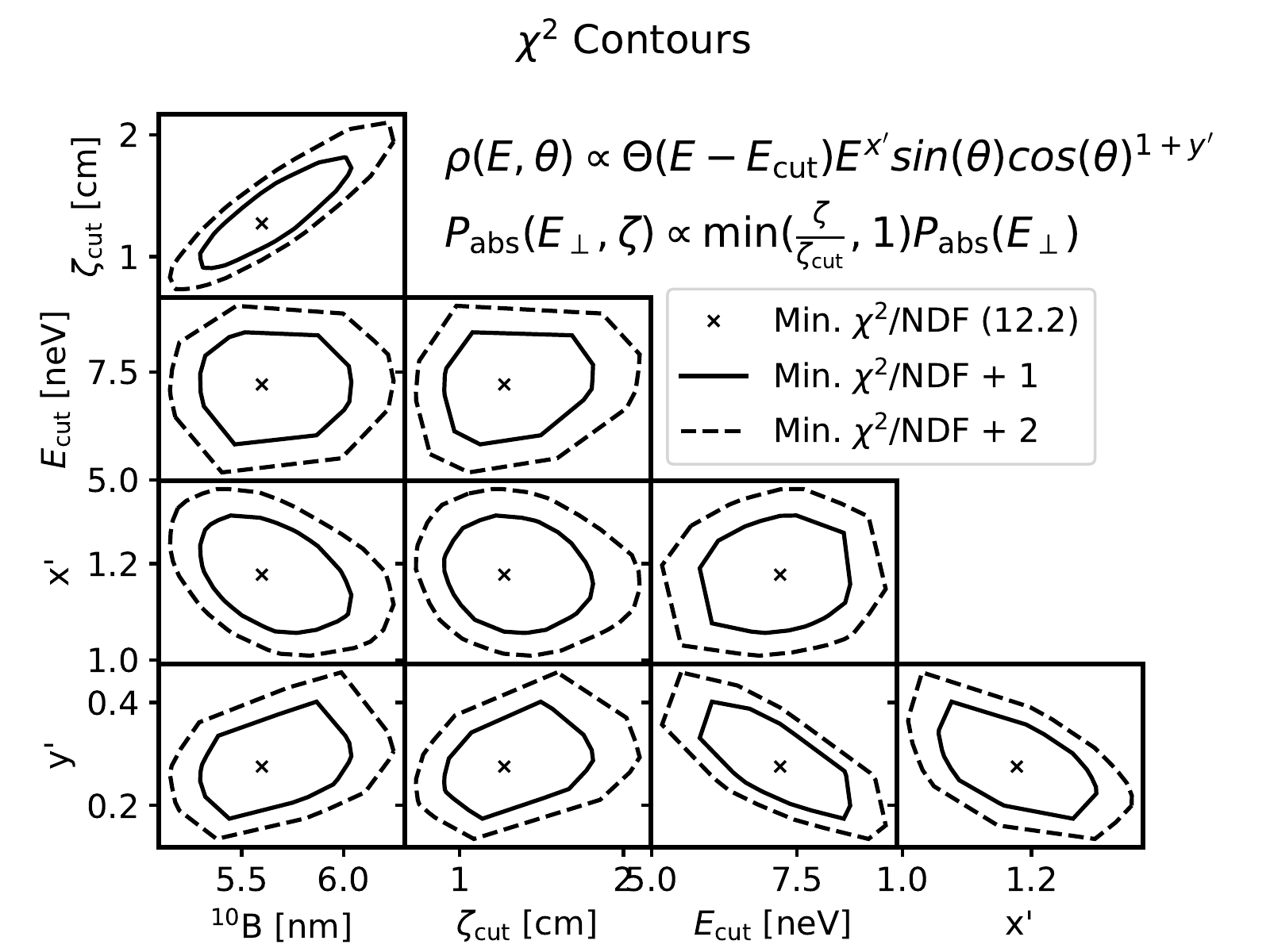}
	\caption{Contours of \(\chi^2\) in the parameter space of \(l_{^{10}\text{B}}\), \(\zeta_{Cut}\), \(E_\text{Cut}\), \(x'\), and \(y'\). On each subplot, the projection of the 5-dimensional \(\chi^2\) contours and the global minimum is shown.}
	\label{fig:chisqcont}
\end{figure*}

In the post-simulation analysis, we construct the timing spectrum of the neutron hit on the dagger detector from simulated data while varying Monte-Carlo parameters. 
We compare the arrival time histogram of the detector hits generated by the Monte-Carlo simulations to the experimental data recorded in the dagger detector for the 2016-2017 run cycle.
Energy-weighted histograms (based on simulations using a superthermal spectrum) can be compared to the reference histograms (based on experimental data) 
to calculate the reduced \(\chi^2\). 
Reference histograms are compiled from dozens of experimental runs under identical conditions but across several weeks. 
Detector counts are formed using the coincidence methods described in previous work \cite{robby}.
Each neutron event is identified by a coincidence of counts from the two PMTs (within a 50~ns time window),
followed by five additional counts where the inter-arrival time was less than 408.8~ns (chosen to minimize the systematic shift due to the detector deadtime/pileup effects).
A \(\chi^2\) value is calculated using the routine in Gagunashvili \cite{gagunashvili} which is also used by the ROOT data analysis framework~\cite{ROOT}.
A \(\chi^2\) minimization was carried out using an optimization routine to determine the model parameters for the best match between the Monte-Carlo data and the experimental data. 
All parameters (\(^{10}\)B thickness, \(\zeta_\text{cut}\), \(E_\text{cut}\), \(x'\), and \(y'\))
were varied to search for the global minimum. 
The $\chi^2$ contours are produced with linear interpolation on results from a subsequent grid search.
2D plots of \(\chi^2\) contours in the parameter space are shown in Fig.~\ref{fig:chisqcont}. Each plot shows the projection of the \(\chi^2\) contours in 2 parameters.

In processing the hit record into the detector timing histogram, we calculate the absorption probability \(\mu(E_\perp)\) using an Akima spline with the GNU Scientific Library~\cite{GNUSciLib}. Instead of calculating Eqs.~\ref{eq:R}, ~\ref{eq:mu}, and ~\ref{eq:absprob}, the spline interpolation speeds up the computation to determine the reflection coefficient for each incident of surface interaction. 
For the typical range of parameters, the difference in \(\chi^2\) between the full model and interpolation was within roundoff error.
The resulting \(\chi^2\) function has many local minima, many of which are artifacts due to discrete steps and statistical fluctuations associated with finite binning size. The Covariance Matrix Adaptation Evolutionary Strategy (CMA-ES) minimization technique was used \cite{hansen}
because it is expected to optimize rugged 
objective functions well.
Global minima were found within hundreds of iterations.
The iteration stops when the solutions converge with a relative difference less than \(\sim10^{-6}\). 
An estimate of the 1-$\sigma$ error in the five fitted parameters is given by forming contours of \(\chi^2_\text{min}+1\). We examine the parameters of several local minima with the lowest \(\chi^2\)s; they all fall within the 1-$\sigma$ range of the best fit parameters.

The dagger detector can be lowered to the lowest point of the trap either in a single step or in multiple steps, each step reaching a subsequently lower position. Multi-step operation allows differential spectral measurement, as only neutrons with high enough energies can reach the detector positioned at elevated heights. As the detector is moved to subsequently lower positions neutrons of decreasing energies are counted in sequence.

\begin{figure}
	\includegraphics[width=0.45\textwidth]{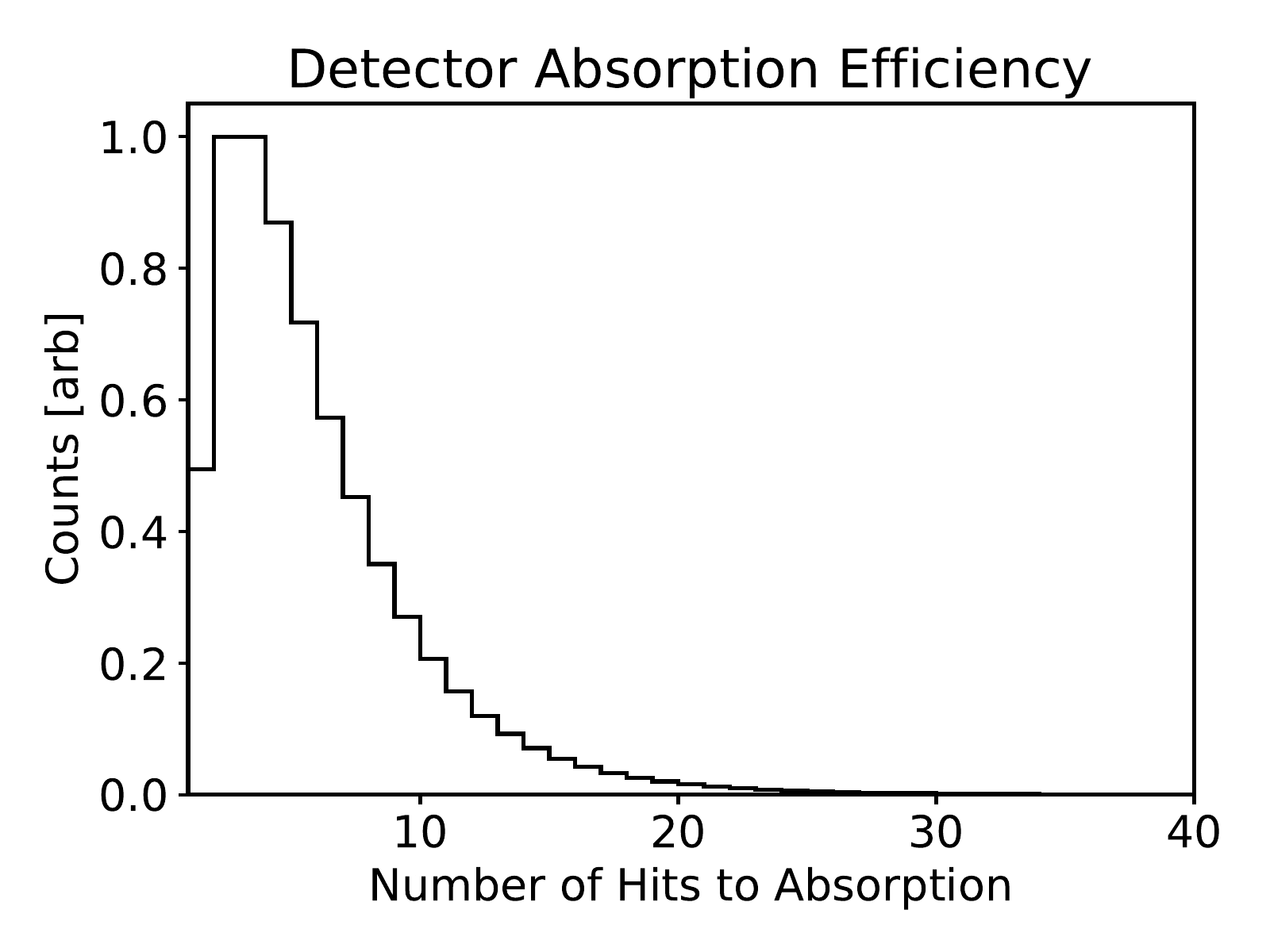}
	\caption{Distribution of the number of collisions until absorption on the detector with a 6~nm $^{10}$B layer. Multiple collisions with the detector took place before a neutron is absorbed by the boron layer and detected.}
	\label{fig:Nhit}
\end{figure}

\begin{figure}
	\includegraphics[width=0.45\textwidth]{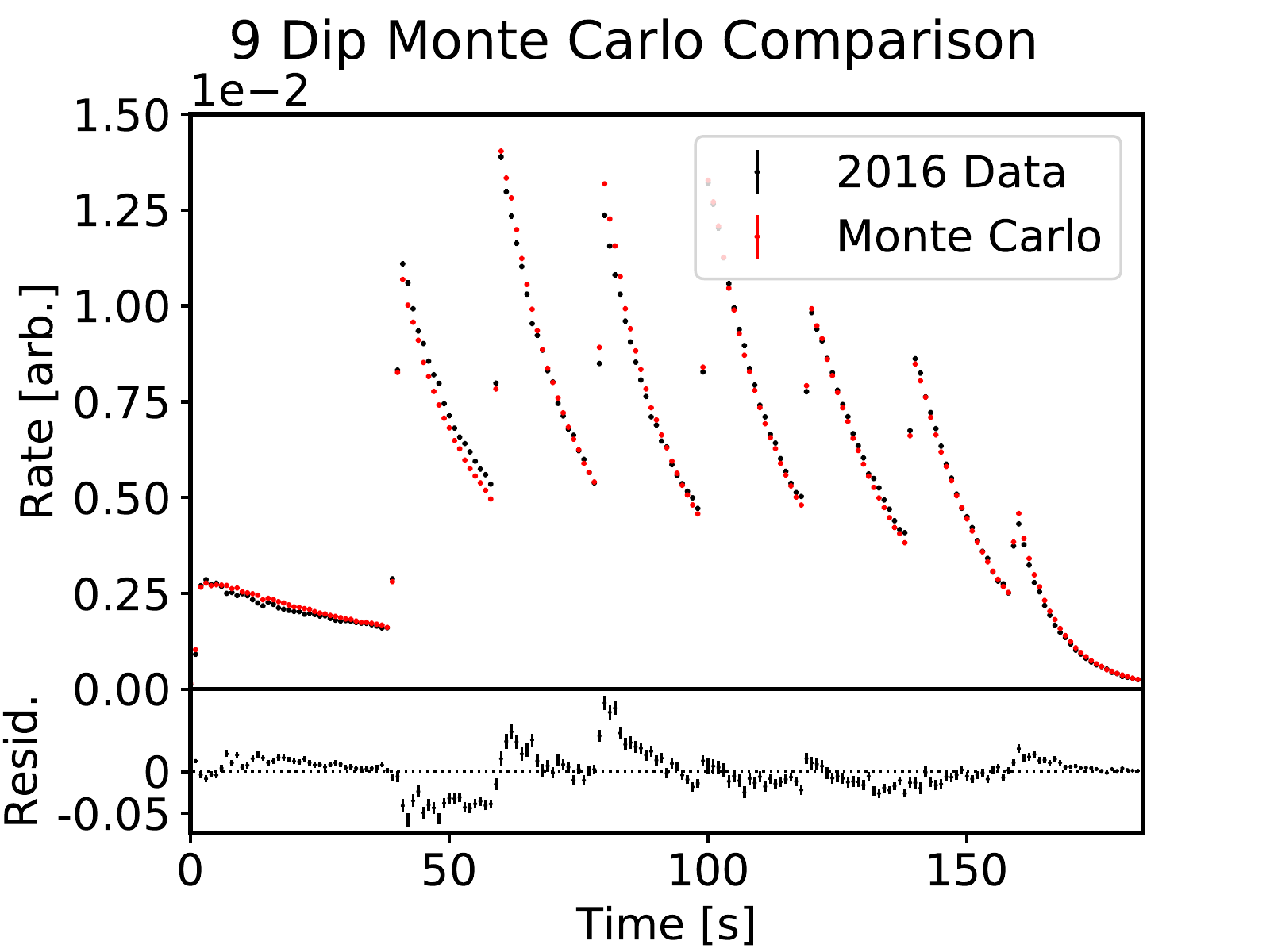}
	\caption{Comparison of 2016-2017 data (9-step measurement) with a simulation with the best-fit Monte Carlo parameters. The absolute difference between the data and the simulation is shown on the bottom figure.}
	\label{fig:9DipComp}
\end{figure}

\begin{figure}
	\includegraphics[width=0.45\textwidth]{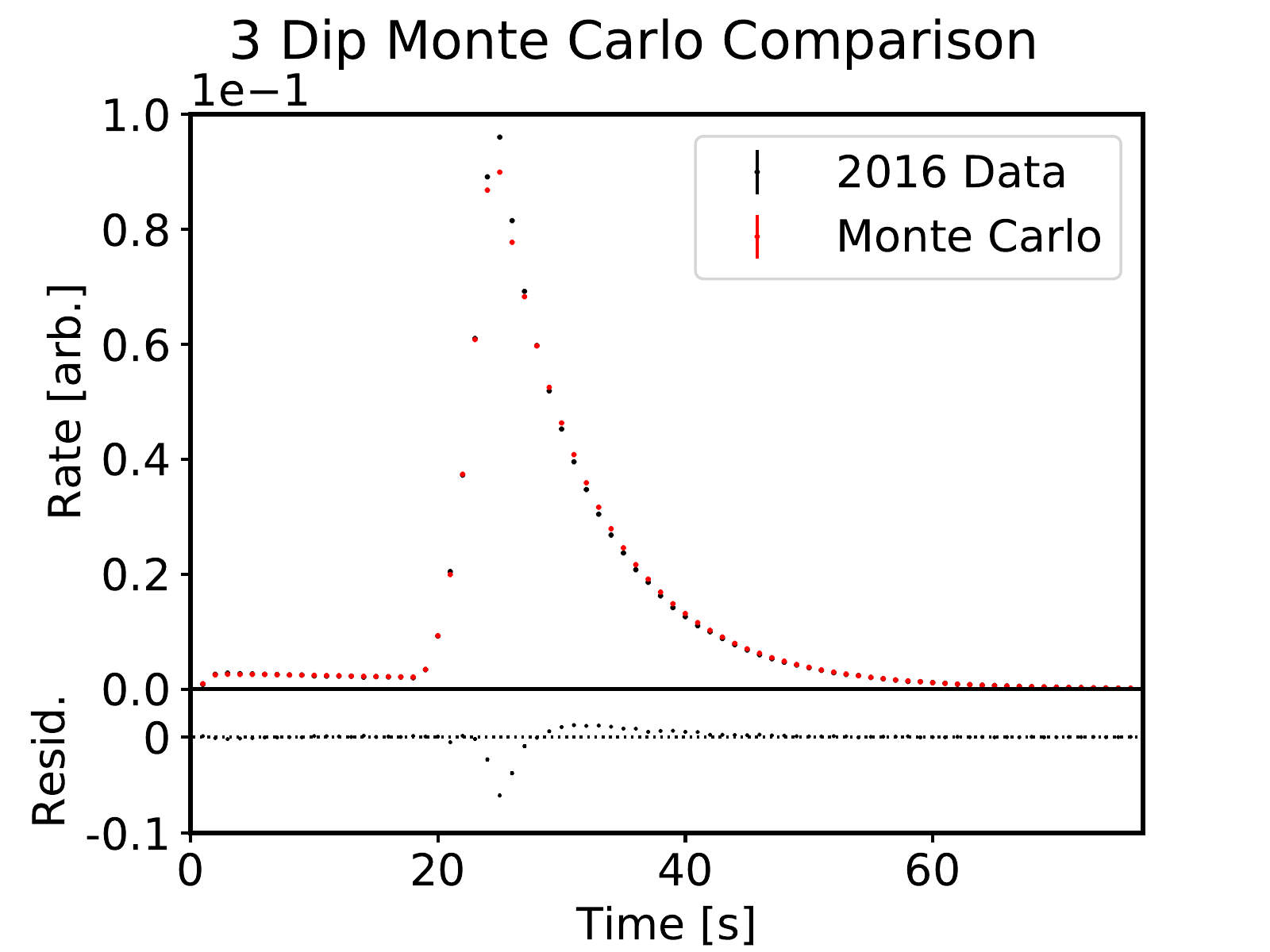}
	\caption{Comparison of the 3-step measurement (2016-2017 data) with a simulation with the best-fit Monte Carlo parameters obtained by the 9-step comparisons (Fig.~\ref{fig:9DipComp}). The absolute difference between the data and the simulation is shown on the bottom figure.}
	\label{fig:3DipComp}
\end{figure}

The model was first tuned on the 9-step data taken in 2016-2017. Recall that the first step of the measurement is meant to detect residual high-energy UCN and is therefore not in the tuning.
A minimum was found with \(\chi^2/\text{NDF}=12\) at \(^{10}\)B thickness = 5.6~nm, \(\zeta_\text{cut}\) = 1.3~cm, \(E_\text{cut}\) = 7~neV, \(x'\) = 1.2, \(y'\) = 0.28.
The \(\chi^2_\text{min}+1\) contours are given in Fig.~\ref{fig:chisqcont} and the histograms are shown in \ref{fig:9DipComp}. 
The analysis favors a detector model with no oxide layer on top of $^{10}$B coating, that is 6~nm, much thinner than expectations, but consistent with the knowledge of the coating thickness of less than 20~nm. We therefore discard the oxide layer in the model.
The thin layer gives the absorption probability per bounce of only 20\%. Distribution of the number of collisions for absorption is plotted in Fig.~\ref{fig:Nhit}.
Neutrons not absorbed scatter off the detector surface as described in Sec.~\ref{sec:ucngen}.
The resulting model with physics of the detector and input spectrum (optimized using the 9-step measurement) was cross-checked by a full simulation to predict the timing spectrum of 3-step measurements.
We also cross-check the energy weighting scheme by comparing histograms of weighted events and histograms of events generated using the optimized model; the results were identical.
Fig.~\ref{fig:3DipComp} shows the results of this optimized simulation in comparison to the 2016-2017 3-step data. 
The resulting \(\chi^2/\text{NDF}\) is comparable to the fit to the 9-step data, indicating consistency.

Even though the simulation with optimized parameters does not sufficiently agree with the experimental data on the quantitative level, as indicated by the rather large $\chi^2$, the exercise brings about qualitative understanding of the neutron arrival time spectra. 
The draining time constant (for each individual peak) is largely controlled by the detector parameters and the relative peak heights by the input neutron spectrum.
On the detector, a larger thickness of the boron layer decreases the draining time observed in all peaks in multi-step measurements. 
Shorter draining times also mean more UCN are counted at the higher heights. 
The energy scaling $x'$ affects the relative height between peaks; 
a higher power scaling will increase the counts in early steps. 
A higher energy cutoff $E_{cut}$ will lower the population of the lowest peaks in the counting period. 
In addition, the angular distribution parameter $y'$ strongly affects the last peak, as UCN with a large initial angle tend to roll on the bottom of the trap and are more likely to be counted when the detector is lowered to the array surface. A more forward-directed incoming flux of UCN decreases the relative height of the last peak.
Finally, the $\zeta_\text{cut}$, which accounts for the damage on the bottom ridge of the detector, adjusts both the relative heights between the peaks and the draining time. A non-zero $\zeta_\text{cut}$ 
allows the bulk of the detector to have a thicker \(^{10}\)B coating, which causes UCN to drain relatively faster in the last step.

\section{Neutron Dynamics} \label{sec:Neutron Dynamics}
With the detector and UCN spectrum model parameters adjusted to produce good  agreement with our measured detector signals,
we attempt to examine the dynamics of the trapped neutrons and to understand whether the procedures of spectral cleaning and neutron detection implemented in the UCN$\tau$ experiment can mitigate systematic effects to the desired level of precision.
The ensemble of trapped neutrons is often treated as 
in an ergodic state which exhibits the same behavior averaged over time as averaged over all the accessible states in the phase space. 
The design and the analysis of past bottle neutron lifetime measurements were largely based on this assumption.
However, this assumption is not valid in our application. Due to the nature of our field-trap, where the neutrons bounce off field gradients of macroscopic scales, it is unlikely that the trapped neutrons would ever establish the ideal stochastic distribution. 
A few special geometries are known to be ergodic, including the 2$d$ stadium and gravitational billiards in a wedge~\cite{chaos}, but most systems exhibit mixed behaviors, with both quasi-periodic and chaotic motions. 
The dynamics of UCN inside our trap influence every stage of the lifetime measurement cycle: it affects how efficiently the overthreshold neutrons are removed in the initial stage of the spectral cleaning; it affects how well the neutrons explore the phase space during the stage of storage; it affects how efficient the neutrons are absorbed and counted by the detector inserted into the trap at different heights.

\subsection{Chaos and UCN Spectral Cleaning} \label{sec:chaos}
The open-top geometry of the UCN$\tau$ trap allows overthreshold neutrons to flow out of the trap. However, many of them remain inside the trap.
If these overthreshold neutrons are in chaotic orbits, then they can quickly find the escape trajectories.
To further facilitate the removal of these overthreshold neutrons, 
a spectral cleaner, made of a large sheet of polyethylene, is inserted to cover a horizontal plane at a fixed cleaning height (typically at a few centimeters below the top of the trap). High-energy neutrons (including the overthreshold neutrons) rise up to the height of the cleaner, get upscattered and removed from the trap. 
However, approximately half of overthreshold neutrons are rolling orbits characterized by trajectories close to the array ($\zeta$ remains within a few centimeters). We call these trajectories rolling orbits because they tend to have their kinetic energy primarily in the longitudinal directions and roll from one end of the array to the other without exploring the inner volume of the trap. As such, they might not rise high enough to interact with the cleaner.
Many of these rolling orbits are quasi-periodic and the time-scale to evolve into escape trajectories, given the non-diffusive nature of the field interactions, can be quite long. During the long storage duration, this slow escape of the residual population of overthreshold neutrons leads to a systematic shortening of the neutron lifetime.

To study the spectral cleaning, we will analyze the dynamics of high-energy neutrons, including overthreshold neutrons.
We classify the motions of these neutrons into the quasi-periodic and chaotic orbits by the Lyapunov Characteristic Exponents (LCE).
The LCE measures the rate of divergence of two trajectories with an infinitesimal difference in the initial condition, 
\(\epsilon\). 
In our 6-dimensional phase space, there are two unique, positive LCE. The largest exponent dominates the growth of the separation \(\vec{\delta}\) because any difference in that direction is exponentially amplified over the other direction.

We follow an algorithm used by Benettin et. al. \cite{Benettin} to measure the largest LCE of any given trajectory.
In our Monte-Carlo simulations, for each trajectory, we randomly sample an initial condition to start a reference trajectory and its partner which is displaced a small amount \(\epsilon\) in phase space. 
The two trajectories, using the symplectic integrator, are evolved for a time \(\Delta t\), followed by an assessment of the separation in phase space \(\vec{\delta}_i\).
The partner trajectory is then reset to begin at a point \(\epsilon\) along \(\hat{\delta}\) away from the reference trajectory.
This procedure is then repeated to allow us to record a distribution of separations.
The record of $N$ separations is averaged to find an averaged rate of separation for each reference trajectory:
\begin{equation}
k=\frac{1}{N\Delta t}\sum_{i=1}^{N}\text{ln}\left(\frac{\lvert\vec{\delta}_i\rvert}{\epsilon}\right).
\end{equation}
The \(k\) approaches the LCE, as \(N\) approaches infinity, and \(\epsilon\) approaches zero.
The normalized separation \(\lvert\vec{\delta}\rvert\) is defined as
\begin{equation}
\lvert\vec{\delta}\rvert = \sqrt{\left(\frac{x-x'}{X}\right)^2 + \left(\frac{y-y'}{Y}\right)^2 + \ldots + \left(\frac{p_z-p'_z}{P}\right)^2},
\end{equation}
where \(x, y \ldots p_z\) is the location of the reference trajectory in phase space, \(x', y' \ldots p'_z\) is the location of the partner trajectory, and \(X, Y \ldots P\) are normalization factors to ensure equal contributions from separations in momentum space and separations in position space.
Both the reference and the partner trajectories share the same initial position and speed, while the initial momenta differ by a slight angular perturbation. This scheme ensures the same energy for the two trajectories.

\begin{figure}
	\includegraphics[width=0.45\textwidth]{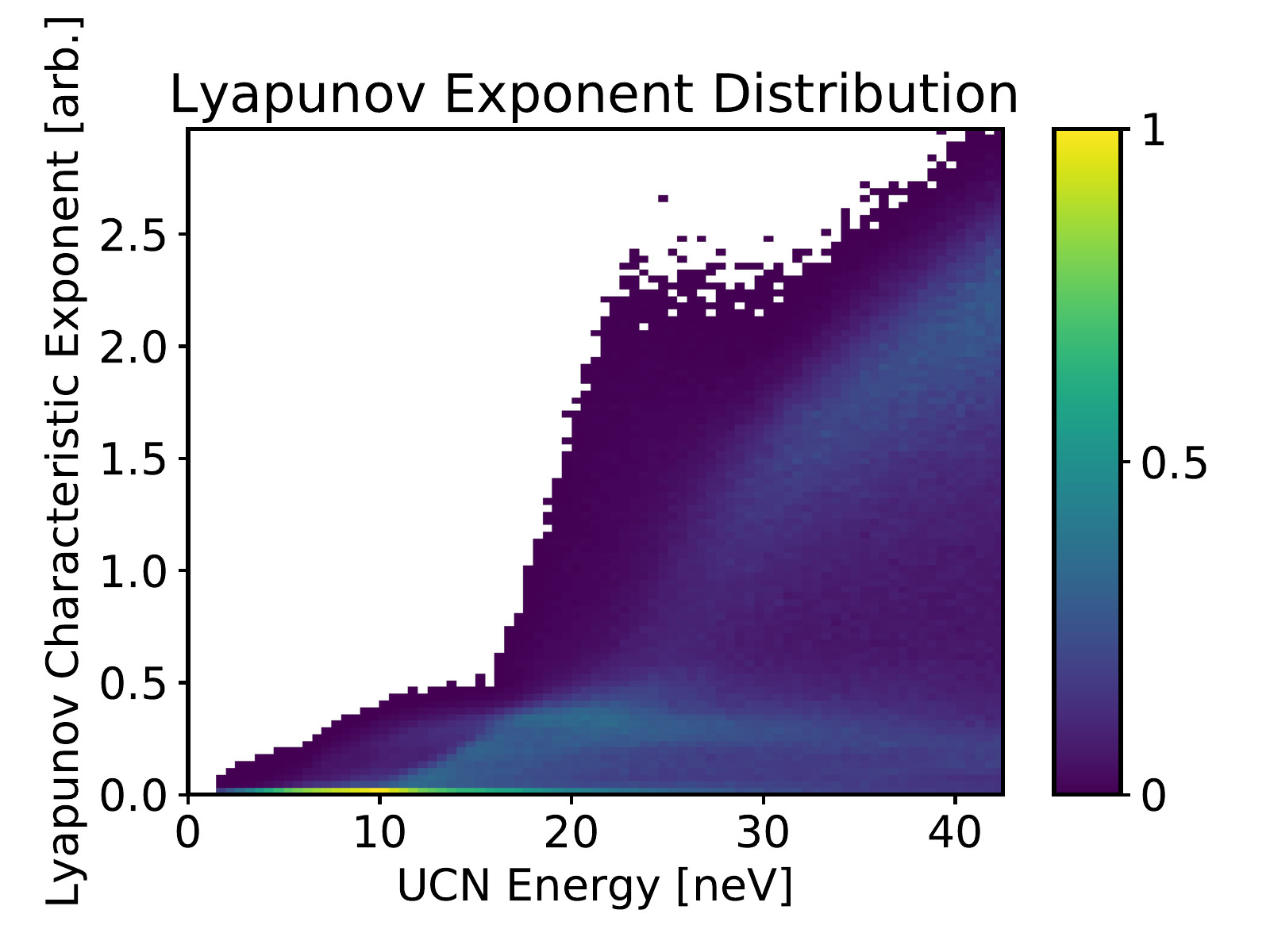}
	\caption{The distribution of LCE of neutron trajectories and the corresponding neutron energy.}
	\label{fig:lyap_by_E}
\end{figure}

Fig.~\ref{fig:lyap_by_E} shows the results of one such Monte-Carlo simulation, in which the trajectories were numerically integrated inside the UCN$\tau$ trap using an initial separation of \(\epsilon=1\times10^{-9}\), for a time \(\Delta t=5\)~s, and for \(N\)=100~resets.
We find that the distribution of the LCE strongly depends on the energy of UCN. Below some critical energy ($E< 25$~neV), most UCN are regular with $k<0.5$. 
Above this threshold, a group of chaotic UCN emerges, with $k$ growing linearly with energy. This increase in $k$ can be understood by a higher momentum causing $\vec{\delta}$ to grow more during the 5~s of separation. As the neutron energy approaches the energy threshold of the trap ($50$~neV), the percentage of neutrons in chaotic orbits increases.
Even for overthreshold neutrons, there is still a non-zero fraction of neutrons in quasi-periodic orbits, which might be hard to remove from the trap and might pose possible challenges in controlling the systematic effects of the lifetime measurements.

\begin{figure}
	\includegraphics[width=0.48\textwidth]{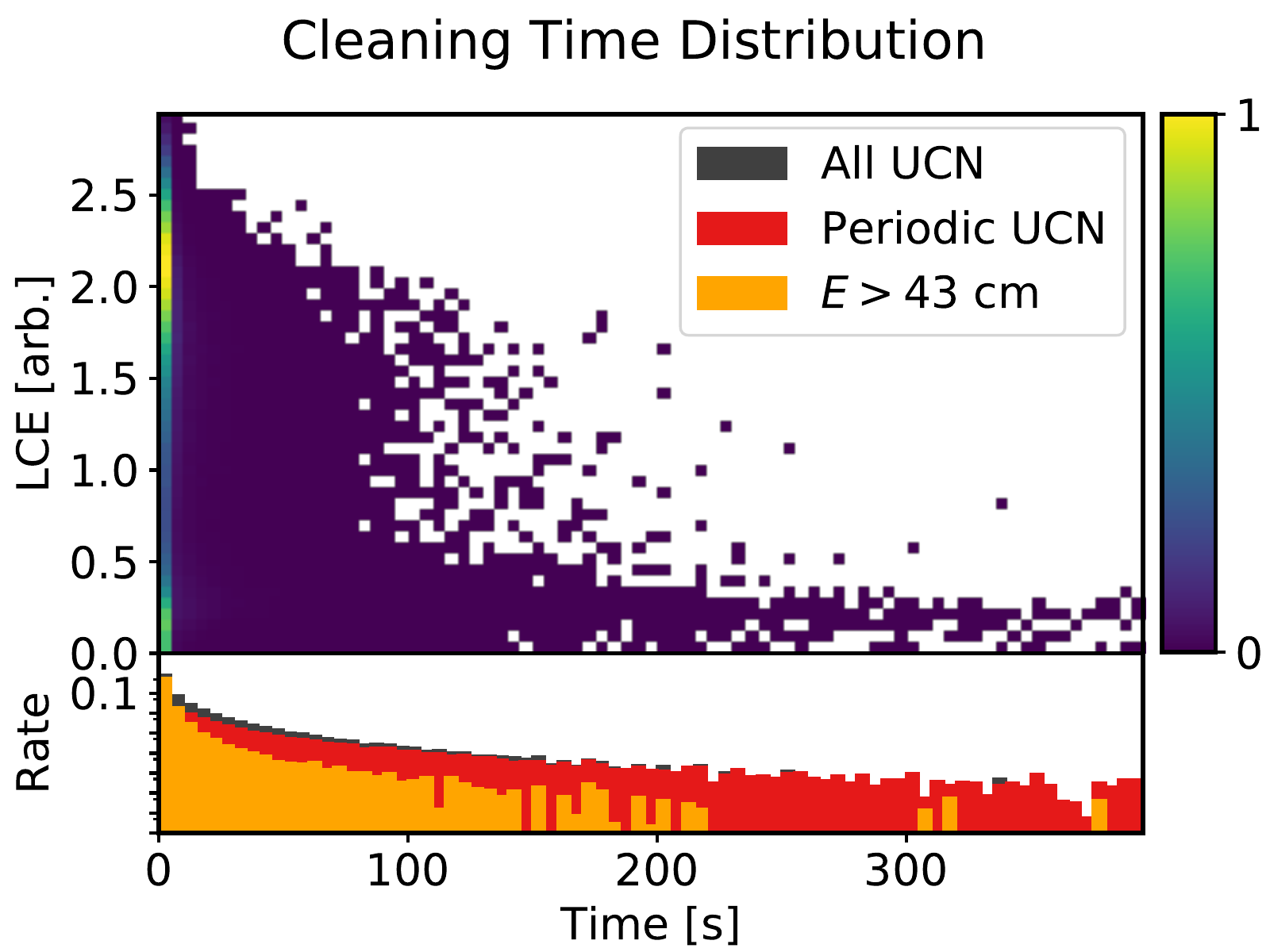}
	\caption{The distribution of LCE and the arrival time on the cleaner. The bottom figure shows the count rate by integrating the neutrons over the full range of the UCN energy (gray), and over neutrons with LCE $<0.75$ with quasi-periodic trajectories (red), and over neutrons with $E/m_n g>43$~cm (orange).}
	\label{fig:lyap_cleaning_time}
\end{figure}

In the recent UCN$\tau$ run cycles, the cleaning height was set at 38~cm above the bottom of the trap. The cleaner covers approximately half of the trap area and was lowered at the start of the filling procedure which lasted 150~s. The cleaner stays at the cleaning height for 50~s, after filling, to remove high-energy neutrons. After the cleaning period, the cleaner was raised to a height of 43~cm, where it remained during the storage period.
In our simulation, the cleaner was modeled as a perfect absorber covering exactly half the area of the trap. UCN were simulated with the cleaner at 38~cm for up to 200~s post-filling. 
The arrival time on the cleaner was recorded for each UCN that could reach it, and we assign the cleaning time as the first arrival time.
Fig.~\ref{fig:lyap_cleaning_time} shows the distribution of the cleaning time for neutrons with high enough energies to rise and intersect the cleaner as a function of time on the $x$-axis. For each cleaner hit event, we also calculate the corresponding Lyapunov exponent $k$, which is displayed on the $y$-axis. 

The 2$d$ histogram, when integrated over the $y$-axis, shows how quickly neutrons are removed from the trap by the cleaner. The cleaning time distribution for all UCN groups is shown in gray. There is an initial fast cleaning period followed by a long tail. The red histogram shows UCN with $k<0.75$. These are trajectories in regular, quasi-periodic orbits, that slowly evolve into trajectories that intersect the cleaner. These quasi-periodic orbits make up the bulk of the cleaning time distribution tail. UCN with energy $>43$~cm are cleaned significantly faster than lower energy UCN. The initial cleaning time is faster and the tail is significantly smaller. These high-energy UCN will be susceptible to losses during storage on the raised cleaner inside the trap.

\begin{figure*}
    \subfloat[Overthreshold neutrons ($E/m_n g> 38$~cm)]{
        \includegraphics[width=0.45\textwidth]{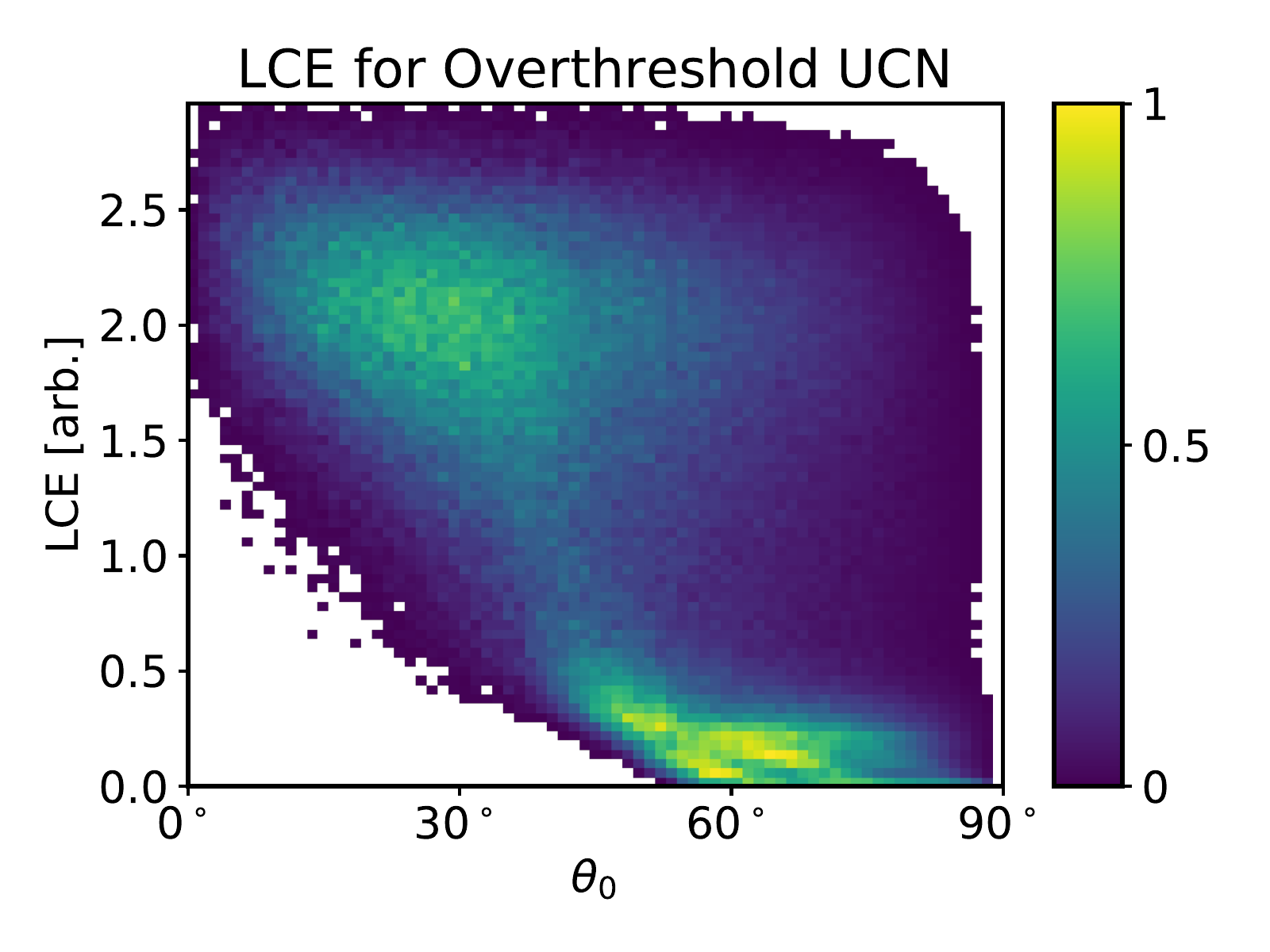}
    }
    \subfloat[Trappable neutrons ($E/m_n g< 38$~cm)]{
        \includegraphics[width=0.45\textwidth]{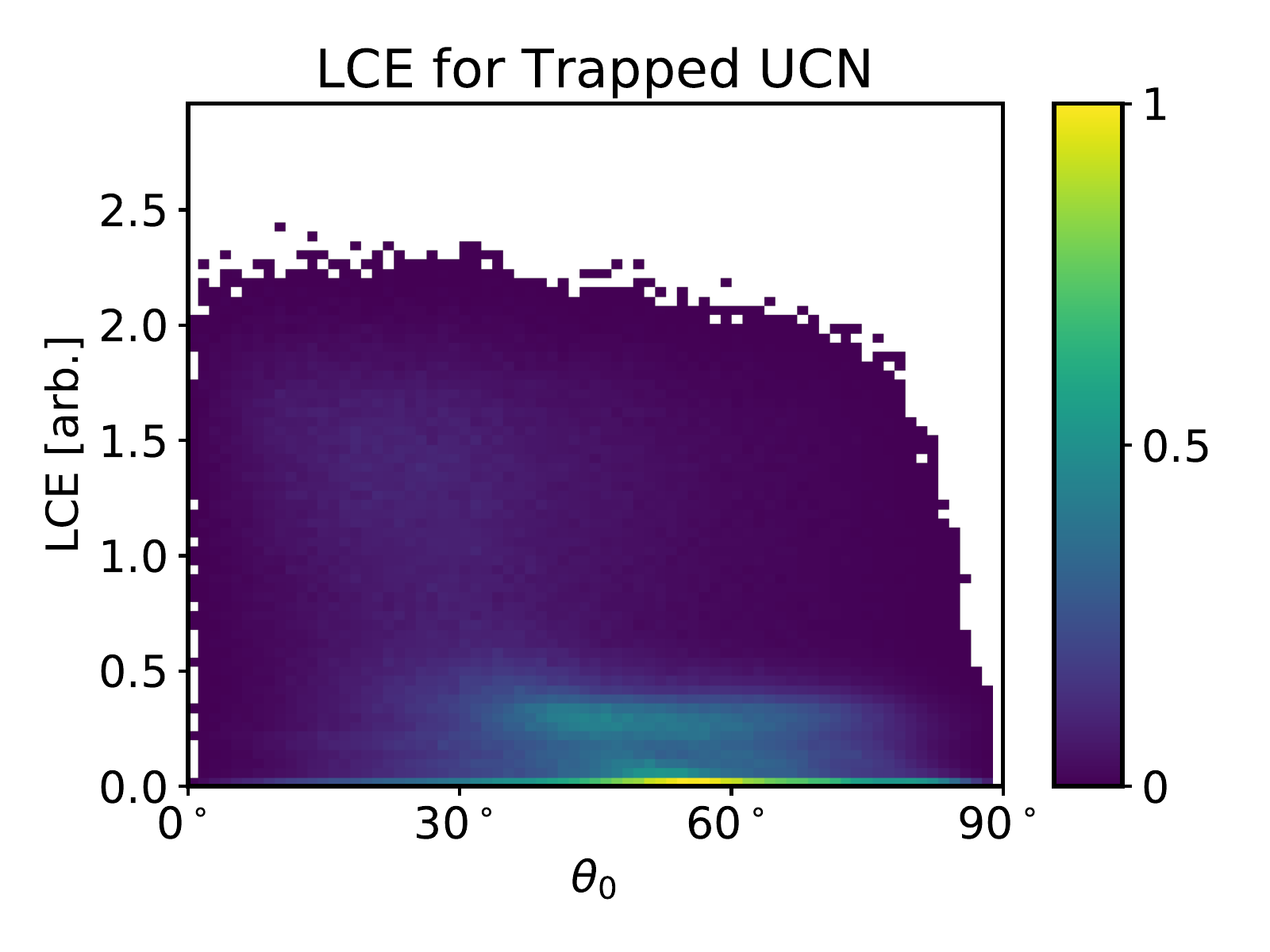}
    }
	\caption{Distribution of the initial polar angle and the Lyapunov exponent for (a) overthreshold neutrons and (b) trappable neutrons.}
	\label{fig:lyap_angle}
\end{figure*}

\begin{figure*}
	\subfloat[\(t<50\)~s\label{fig:cleaner_hit_short_t}]{
        \includegraphics[width=0.45\textwidth]{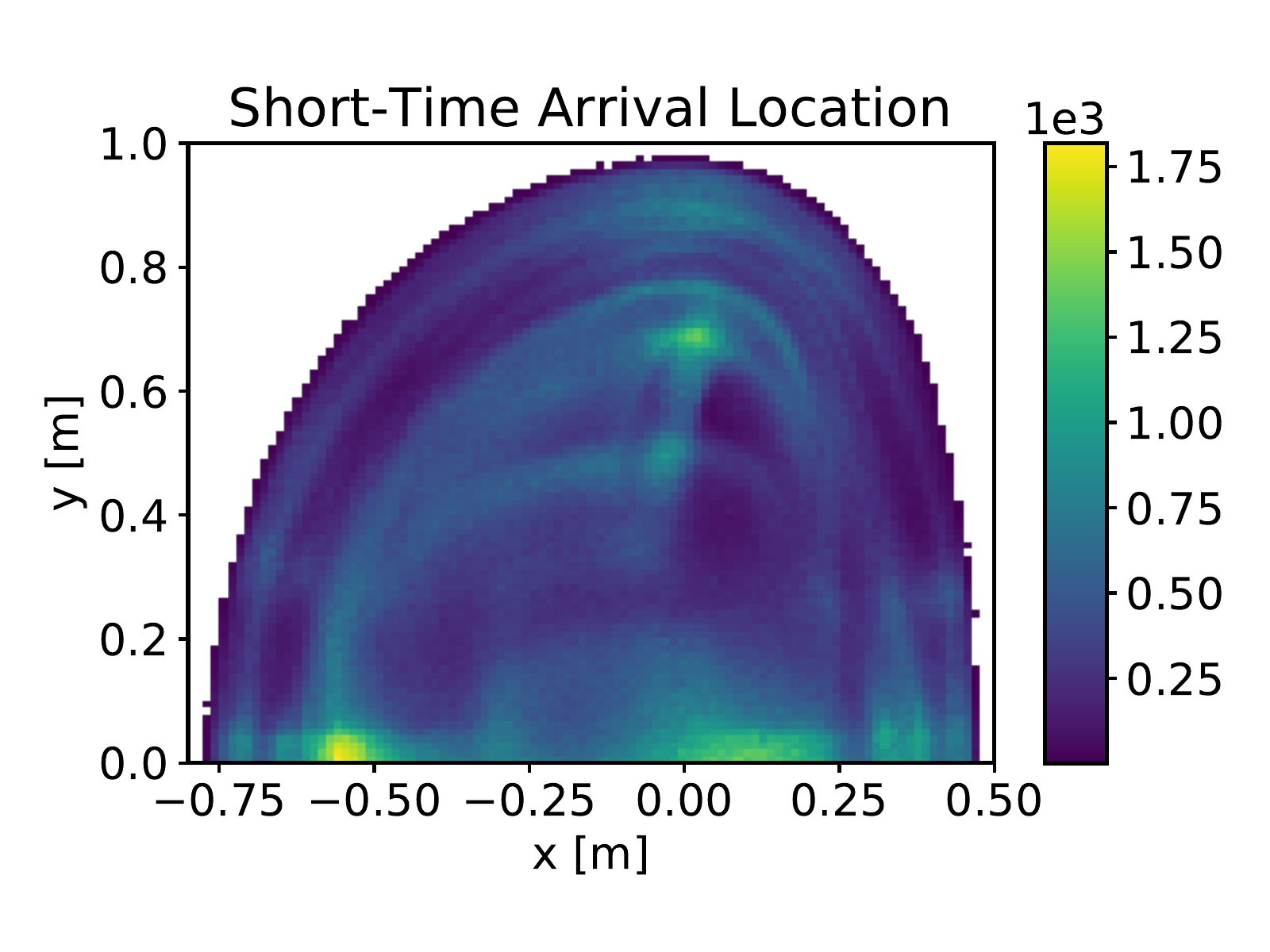}
    }
	\subfloat[\(t>50\)~s\label{fig:cleaner_hit_long_t}]{
        \includegraphics[width=0.45\textwidth]{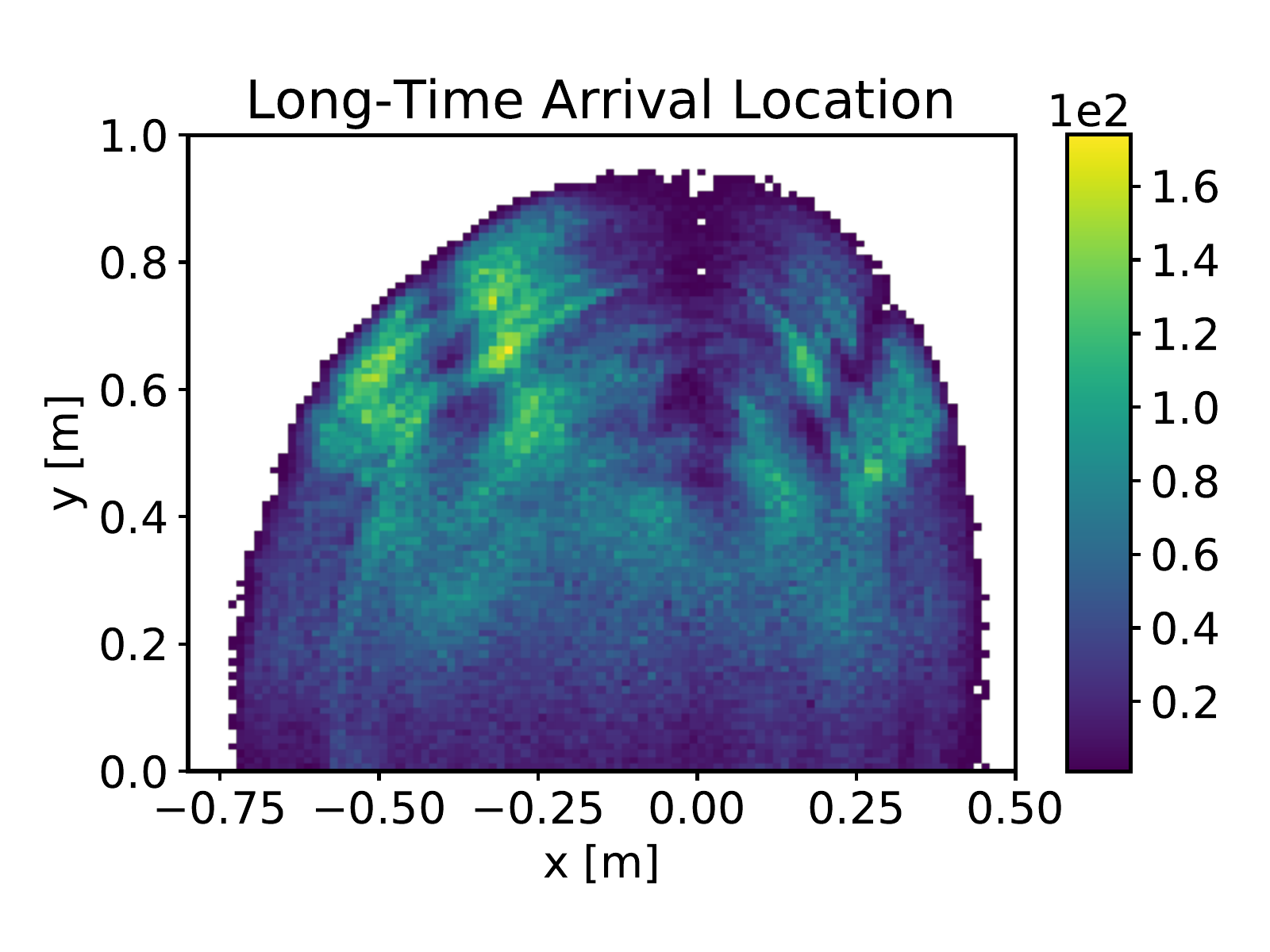}
    } \\
    \subfloat[$\theta_0>45^\circ$ \label{fig:cleaner_hit_high_theta}]{
        \includegraphics[width=0.45\textwidth]{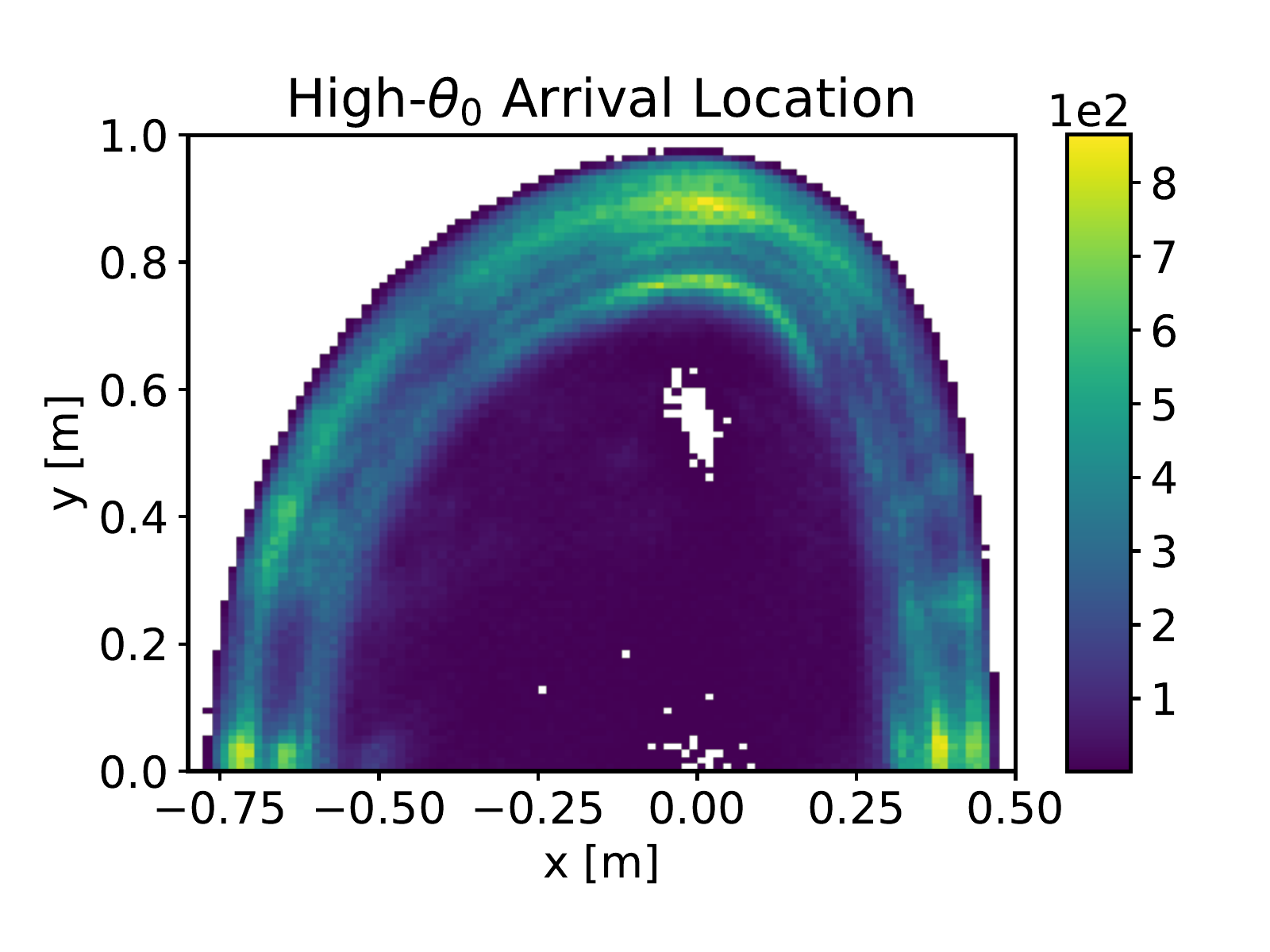}
    }
	\subfloat[Heated neutrons on Raised Cleaner\label{fig:cleaner_hit_raised}]{
        \includegraphics[width=0.45\textwidth]{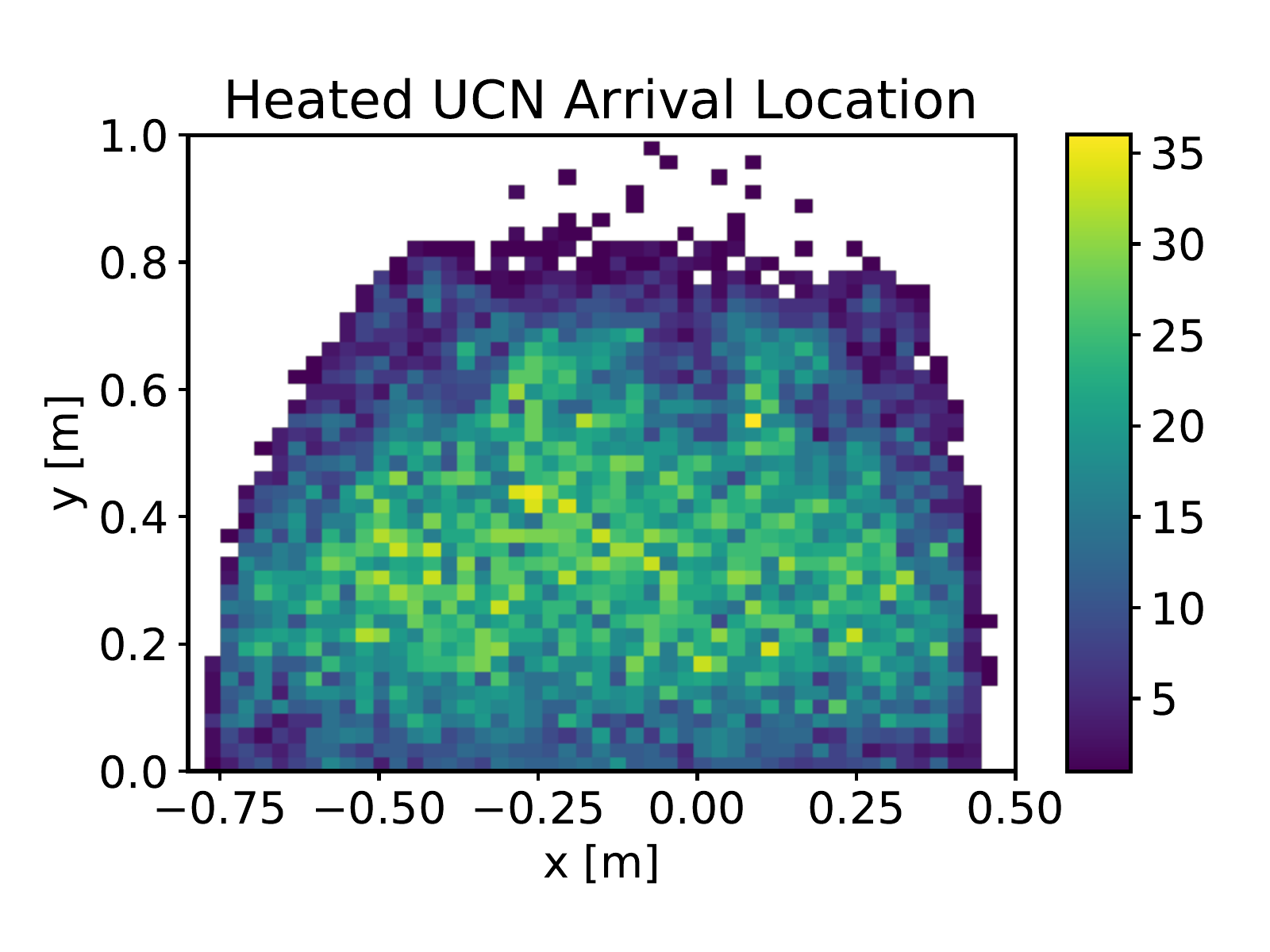}
    }
	\caption{Distribution of neutron hits on the cleaner surface for (a) neutrons arriving over the first 50~s of the cleaner insertion, (b) neutrons arriving after 50~s, (c) neutrons with initial polar angle $\theta_0>45^\circ$ (integrated over the full cleaning time), and (d) neutrons being heated during storage and hitting the cleaner at the raised position.  }
	\label{fig:cleaner_hit}
\end{figure*}

To understand how orbits of different LCE are populated, we plot the initial polar angle $\theta_0$ and the Lyapunov exponent of the resulting trajectory. Fig.~\ref{fig:lyap_angle} shows the distribution for both the overthreshold neutrons with $E>g \cdot 38$~cm and the trappable neutrons with $E<g \cdot 38$~cm.
For overthreshold neutrons (on the left plot), the neutrons are roughly separated into two groups by the LCE: the group with $k>0.75$ has a broad angular distribution centering at low $\theta_0$ and the other with $k<0.75$ starts with larger $\theta_0> 45^{\circ}$.
A sizable amount of overthreshold neutrons are quasi-periodic and in rolling orbits, populated with an initial polar angle $> 45^{\circ}$. 
These neutrons, as shown in Fig.~\ref{fig:lyap_cleaning_time}, take a long time to clean due to their small LCE. In contrast, Fig.~\ref{fig:lyap_angle}b shows that the trappable neutrons, independent of their initial polar angle, mostly reside in quasi-periodic orbits with small $k$.
To understand how to most efficiently clean these neutrons,
we plot the hit position on the cleaner in Fig.~\ref{fig:cleaner_hit_short_t}. 
The non-uniform distribution of the hit position can be understood as the effects of imaging the trapdoor (where the neutrons are started) by the combined magnetic and gravitational fields; the effective neutron optics renders on the plane of the cleaner multiple images of the trapdoor, similar to the effects of gravitational lensing. 
Fig.~\ref{fig:cleaner_hit_high_theta} shows how the hit position of neutrons in the rolling orbits (neutrons with $\theta_0>45^\circ$) concentrate on the edge of the cleaner. Fig.~\ref{fig:cleaner_hit_long_t} shows the hit of neutrons after the cleaning time of 50~s; the resulting uncontrolled removal of neutrons will cause a systematic shift on the neutron lifetime. Finally, Fig.~\ref{fig:cleaner_hit_raised} shows the neutron hit distribution if the neutrons are heated during the storage; these neutrons are lost on the cleaner during the storage time and it leads to another systematic shift in the neutron lifetime.

To reduce the size of the systematic lifetime shift due to insufficient spectral cleaning, we can either leave the cleaner at the cleaning height (38~cm) for longer times, or move the cleaner at a lower height while maintaining the same raised height. The resulting statistics of neutrons in these cleaning scenarios are given in Tab.~\ref{tab:clean}. The numbers are reported as the ratio to the total number of neutrons.
In this table, the uncleaned neutrons are defined as the total neutrons with sufficient energy to reach height larger than the cleaning height of the cleaner (38~cm). Out of these, some can reach higher than 43~cm, which is the height of the cleaner at the raised position. These high-energy neutrons have a finite probability to hit the cleaner at the raised height and get lost during storage, causing a systematic shortening of the measured neutron lifetime.
This leads to a neutron lifetime down-shift of 0.034 second, which is consistent with the upper-bound limit reported in Ref.~\cite{robby}.

\begin{table}
  \begin{ruledtabular}
  \begin{tabular}{l r r r}
      Cleaning Conditions & \multicolumn{2}{c}{Population Fraction} & \(\Delta\tau\) [s]\\
      time, height, prob.& $E/m_n g>38$~cm & $> 43$~cm & \\
      \colrule
      50~s, 38~cm, 50\% & 0.024 & 9.8\(\times10^{-5}\) & 0.05\\
      50~s, 38~cm, 100\% & 0.012 & 6.6\(\times10^{-5}\) & 0.034\\
      200~s, 38~cm, 100\% & 0.0036 & 3.4\(\times10^{-6}\) & 0.0016\\
      50~s, 35~cm, 100\% & 0.00019 & 9.8\(\times10^{-8}\) & 8\(\times10^{-5}\)\\
  \end{tabular}
  \end{ruledtabular}
  \caption{Fraction of uncleaned UCN population and the shift in \(\tau\), evaluated with cleaning configurations varied in the cleaning time, the cleaning height, and the ubscattering probability per interaction. The high-energy cut-offs of interests are set by the cleaner at the cleaning position at 38~cm and at the raised position at 43~cm. }
  \label{tab:clean}
\end{table}

\subsection{Multistep Detection \& Peak 1 Corrections} \label{sec:p1corr}

The dagger detector provides insight into the dynamics of the trapped UCN, and understanding the detector response using both Monte Carlo and data-driven methods is critical for limiting potential systematic effects.
Multi-step neutron detection spreads out the neutron counts over time and thus mitigates the size of dead-time related corrections associated with high counting rates. More importantly, it allows for differential spectral measurement, as only neutrons with high enough energies can reach the detector positioned at elevated heights.
In a typical multi-step detection, 
the first step, set at the cleaning height (38~cm), can be used to check for the presence of uncleaned or heated neutrons. The arrival time profile of neutrons during the first step is referred to as Peak 1 or P1.
The shape of the timing spectra measured in the multi-step detections is a product of the geometrical acceptance and the spectral distribution of the stored neutrons. The resulting spectral monitoring is a handle to detect systematic effects, such as material loss or heating loss, that often have a strong energy dependence.

\begin{figure}
	\includegraphics[width=0.45\textwidth]{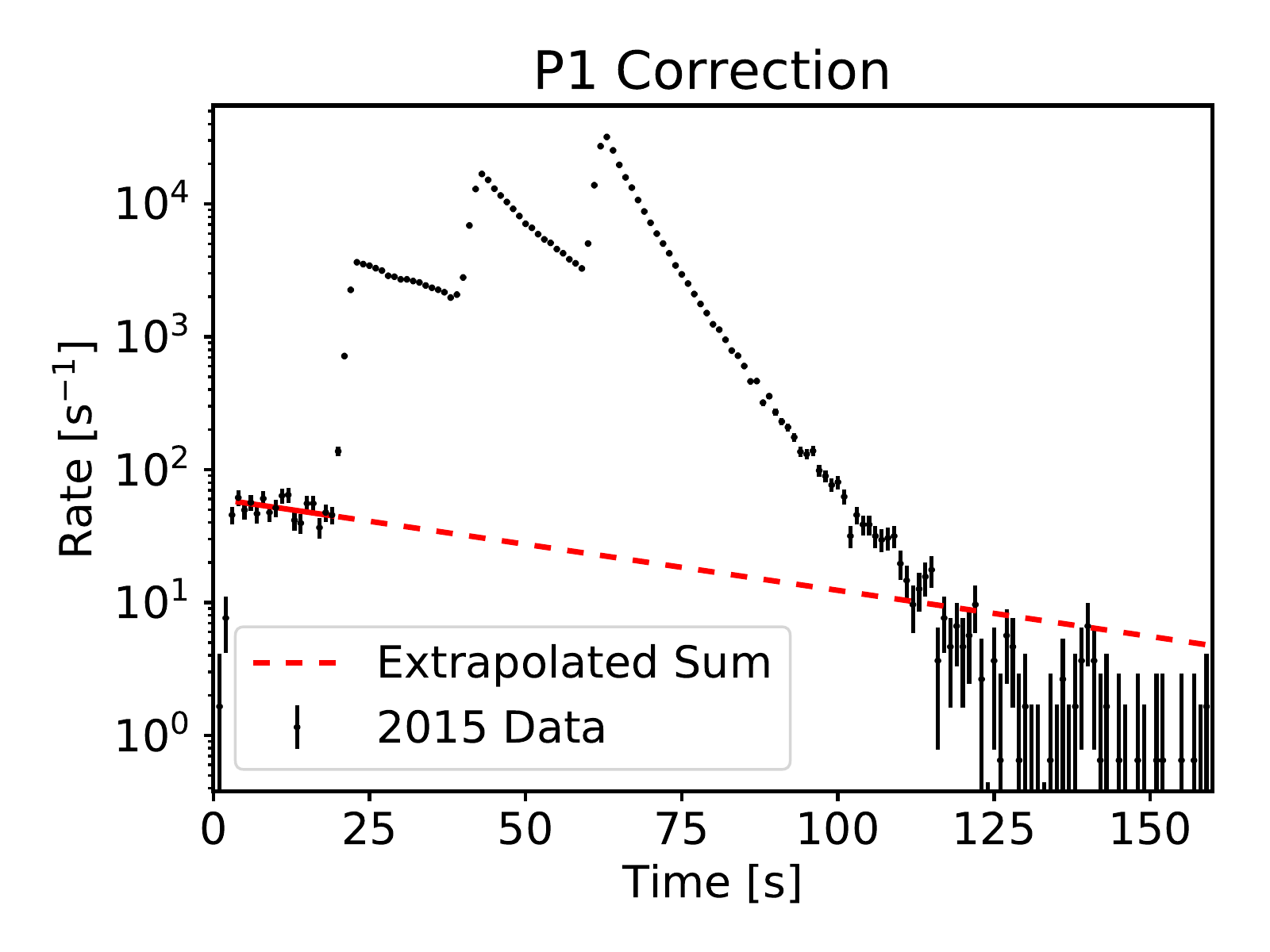}
	\caption{The P1 subtraction procedure is illustrated by the red dashed curve which fits the neutron counts measured in P1 (as the detector is positioned in the cleaning height) by an exponentially decaying curve, extrapolated to infinite time. The curve fitting estimates the total population of overthreshold neutrons that did not get removed by the cleaner; the fitted number is then subtracted from the total counted neutron events.}
	\label{fig:P1subtraction}
\end{figure}

We have attempted to use the neutrons counted in P1 to quantify the effects of insufficient cleaning.
If the cleaner did not sufficiently remove the overthreshold neutrons (as discussed in the previous section), then we expect a finite probability for these overthreshold neutrons to be counted by the dagger detector placed at the cleaning height. We expect these excess counts for the short storage runs, but not for the long storage runs because during the long storage time the overthreshold neutrons would have either exited the trap or intersected with the cleaner at the elevated height and thus evaded detection. 
This leads to a systematic downshift of the measured neutron lifetime.
To correct for this effect of insufficient spectral cleaning, we implemented a correction procedure, called P1 subtraction. 
We subtract the residual population of overthreshold neutrons from the total detected counts, based on the counts registered in Peak 1 of the multi-step detection when the detector was inserted at the cleaning height.   
The geometrical acceptance of neutrons for the dagger detector scales roughly as the area of the active surface of the detector to the volume occupied by the neutrons (of sufficient energies to intersect the dagger detector at the specific height). Thus, the geometrical acceptance in the P1 position is rather limited, as indicated by the long counting time recorded in peak 1 in the multi-step timing spectrum (see Fig.~\ref{fig:P1subtraction}).
In our P1 correction procedure, we extrapolate the total population of the uncleaned, overthreshold neutrons in the first peak by fitting the counting curve to an exponential curve (the red dashed line in Fig.~\ref{fig:P1subtraction}) with the draining time constants measured in dedicated runs. Integrating this extrapolated curve to infinite time should give the total number uncleaned UCN which are overthreshold and could rise up the the height of the cleaner. See Ref.~\cite{morrisrsi,robby} for more details.

The correction procedure described above assumes that all the overthreshold neutrons can be counted by the dagger detector in the P1 position, if the detector were counting in position for a sufficiently long time. 
The curve-fitting extrapolates the integrated number of overthreshold neutrons and thus corrects for the finite measurement time. 
It implicitly assumes that the geometrical acceptance of overthreshold neutrons remains the same independent of the detector position and height; Naively, the acceptance is roughly the ratio of the area of the detector to the volume occupied by the overthreshold neutrons.
These assumptions are valid, if the neutron distribution is ergodic, i.e., the trapped neutrons sample uniformly through the accessible points in the phase space volume.

\begin{figure}
	\includegraphics[width=0.45\textwidth]{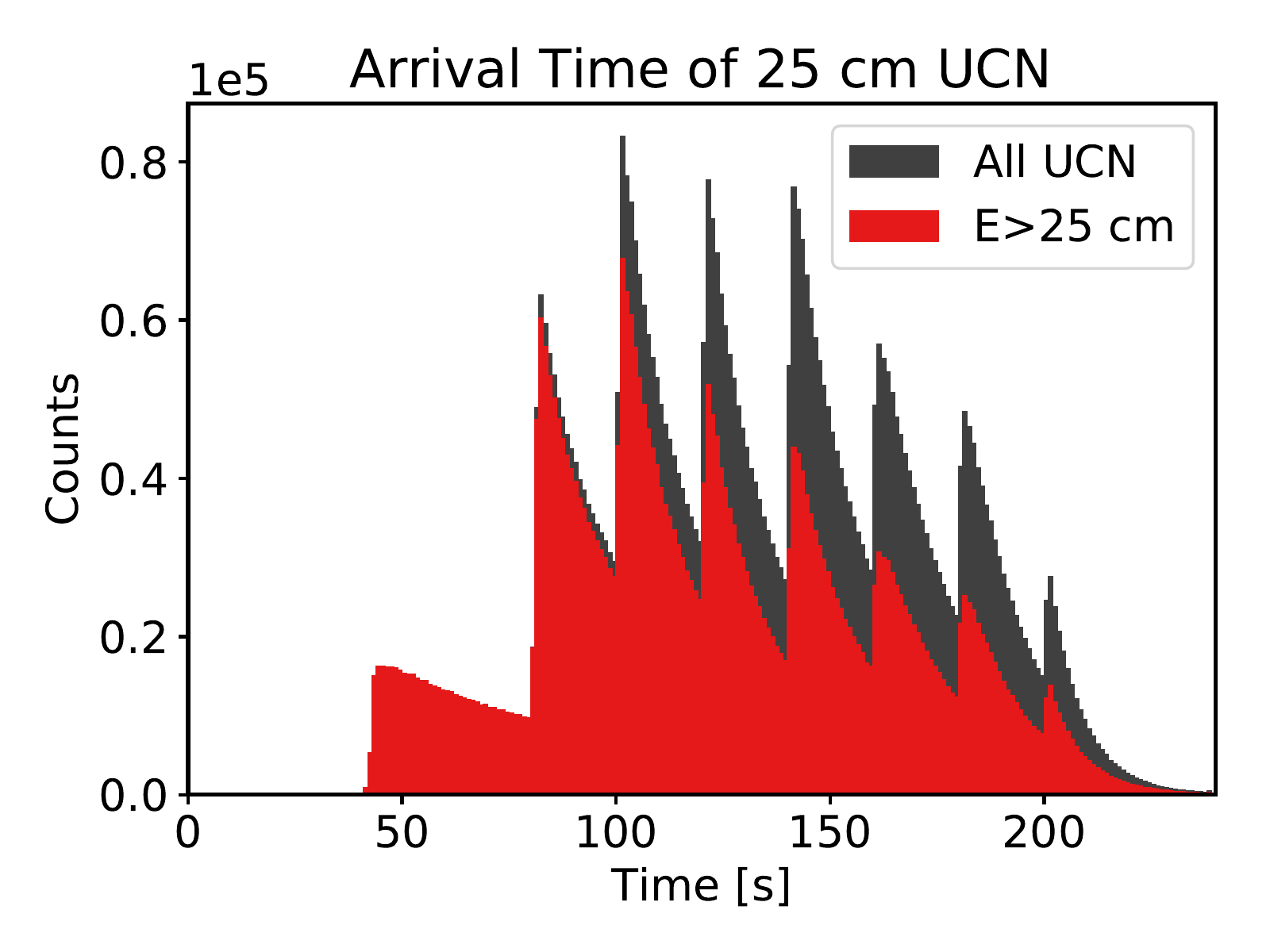}
	\caption{The detection of neutrons (all energies) in the 9-step detection (the gray histogram) is compared to the 
detection of high-energy neutrons, $E/m_ng> 25$~cm (the red histogram). The first peak has no counts due to the high efficiency of the cleaner. The second peak measures neutrons with $E/m_ng> 25$~cm. Note that the subsequent peaks at increasingly lower height contains sizable population of the same high energy neutrons with $E/m_ng> 25$~cm. The extrapolated sum of P2 counts only 30\% of the total UCN with $E/m_ng> 25$~cm.}
	\label{fig:HighE}
\end{figure}

This simulation work, on the other hand, shows that a significant amount of field-trapped neutrons remain in quasi-periodic trajectories and they don't fully explore the phase space. 
In the UCN$\tau$ trap, there are many neutrons residing in quasi-periodic orbits, characterized by small LCEs. The simulation shows that after the spectral cleaning (50~s), there remains a significant population of overthreshold neutrons in quasi-periodic orbits, which were initially populated with large polar angles (see discussions in the previous section). These neutrons bounce close to the surface of the trap and rise up in height, only when they come close to the circumference of the trap. 
Thus, the geometrical acceptance of these overthreshold neutrons by the dagger detector placed in P1---high in the middle plane of the trap---is practically zero.
As the dagger detector descends to the subsequently lower position it overlaps with the volumes occupied by overthreshold neutrons,  and the geometrical acceptance increases.
This effect is illustrated in the timing spectrum shown in Fig.~\ref{fig:HighE}, in which the 9-step detection of the full UCN spectrum (the gray histogram) is compared to the detection of high-energy neutrons with $E/m_n g> 25$~cm (the red histogram).
As the detector descends to a lower position, more high-energy neutrons are counted; the total number of neutron counts is 3 times greater than the extrapolated exponential decay assumed in the P1 subtraction procedure.
The P1 subtraction procedure thus (1) has little sensitivity to populations of overthreshold neutrons with low LCEs, and (2) uses the exponential function that fails to capture the geometrical acceptance of overthreshold neutrons for subsequent lower steps.
Limits for the residual population of overthreshold neutrons from our P1 subtraction procedure are difficult to interpret, in that the sensitivity to these populations appear to be low when measured in P1 and the extrapolation procedure fails to account for acceptance of higher energy UCN in the trap in subsequent, lower steps.  
Because our corrections are sensitive to populations of UCN which are expected to be small, evolve very slowly in phase space, and may have low detection efficiency, providing an reliable Monte Carlo prediction for the sensitivity of the P1 subtraction scheme is under development. In particular, we are in the process of refining both our simulation and measurements which explicitly test our ability to model phase space evolution.

\begin{figure}
  	\includegraphics[width=0.45\textwidth]{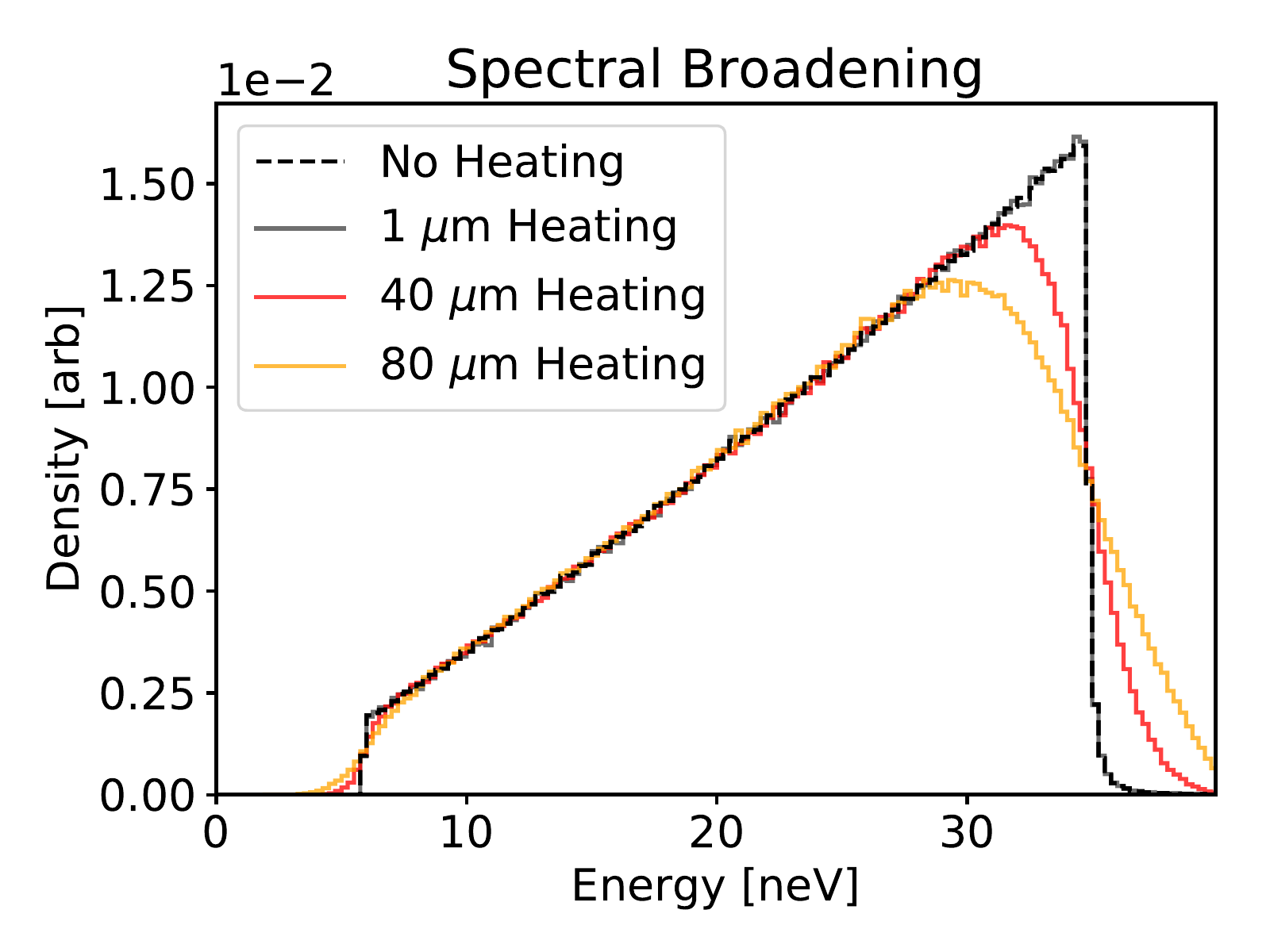}
    \caption{The spectrum of detected neutrons in the simulation with vibration of microphonic frequencies. Scenarios of no vibration, 1~\(\mu\)m, 40~\(\mu\)m, and 80~\(\mu\)m are shown. The neutron spectrum broadens with increasing amplitude of vibration.}
    \label{fig:spectbroad}
\end{figure}

Our simulations do indicate, however, that the systematic shift in the neutron lifetime, due to insufficient cleaning, is well under control as long as the large cleaner is used. Thus 
the P1 subtraction procedure does not have direct consequence to the neutron lifetime.
The same arguments also apply to the way we quantify the effect of neutron heating. 
UCN could be heated due to mechanical vibrations of the apparatus.
The effects of microphonic vibrations can be modeled by adding oscillatory time dependences in the coordinates used to describe the magnetic field. 
We have implemented a more comprehensive heating models in our simulations, 
building upon a 1D toy model used previously~\cite{salvat}.
Vibrations of the Halbach array were modeled by adjusting the $x$-$y$-$z$ coordinates of the array by \(\delta x(t), \delta y(t), \delta z(t)\). We use realistic vibrational amplitudes \(\delta x(t), \delta y(t), \delta z(t)\) extracted from accelerometer measurements, which shows multiple frequencies below 200~Hz and amplitudes no larger than 1~\(\mu\)m. 
For every field bounce, neutrons in the trap can gain or lose energy. After a sufficient number of bounces, the energy distribution broadens; neutrons on the high-energy end of the distribution can rise above the cleaning height. Fig.~\ref{fig:spectbroad} shows the effects of spectral broadening due to vibrational energy transfer. Over time, a small number of neutrons gain enough energy to reach the raised cleaner and get absorbed. Fig.~\ref{fig:cleaner_hit_raised} shows the hits of heated neutrons in the raised cleaner.
The simulation records the number of neutrons lost on the raised cleaner due to vibrations to determine the shift in lifetime. 
Using the vibration data, we conclude that the loss due to microphonic heating can downshift the neutron lifetime by \(\delta \tau=0.03\)~s. This shift is dominated by initially uncleaned neutrons and is consistent with shifts due to insufficient cleaning.

\begin{figure*}
    \subfloat[The detection of uncleaned neutrons in a short-storage run. \label{fig:p1effUncleaned}]{
        \includegraphics[width=0.45\textwidth]{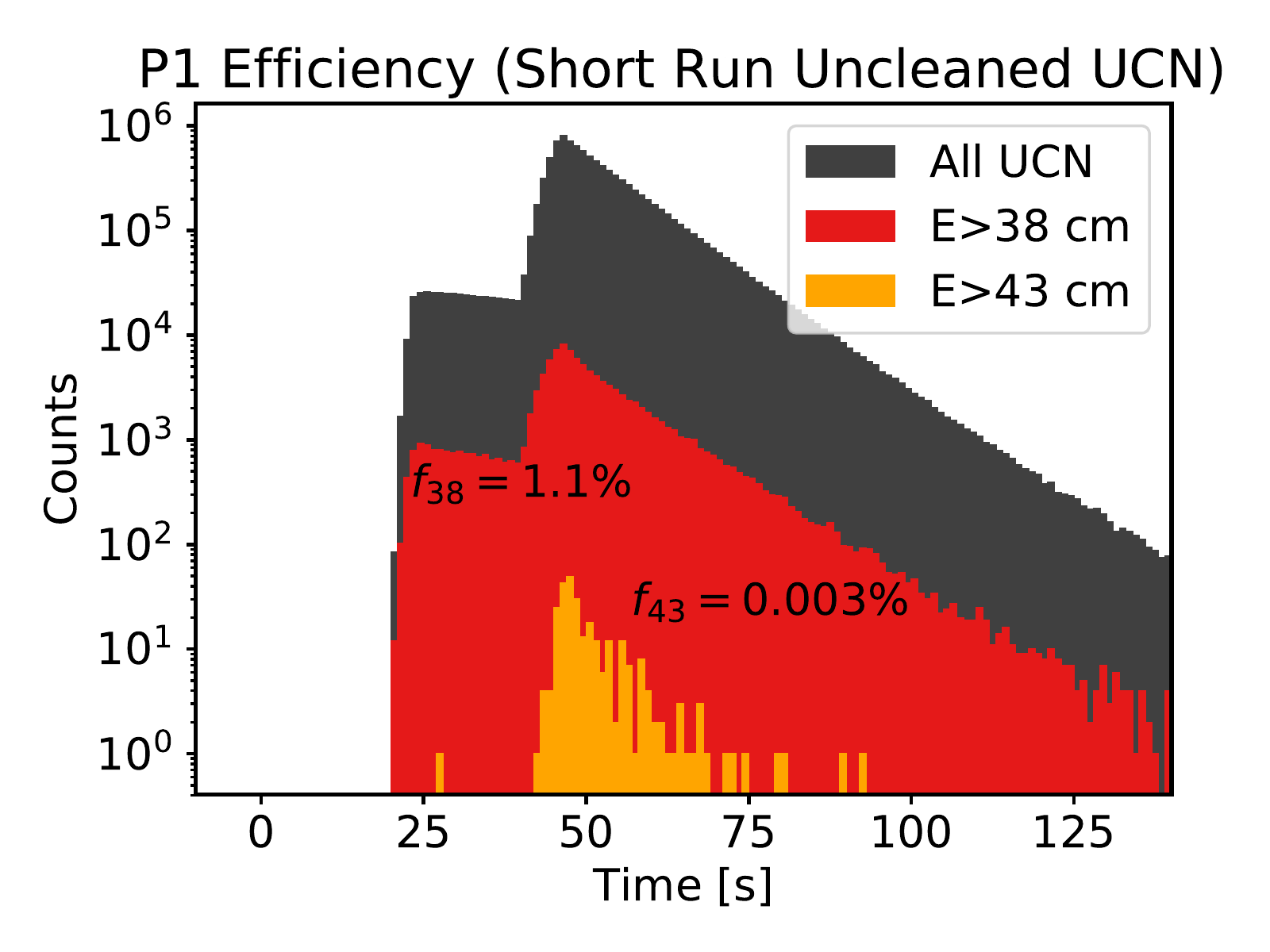}
    }
    \subfloat[The detection of heated neutrons in a long-storage run.\label{fig:p1effHeated}]{
        \includegraphics[width=0.45\textwidth]{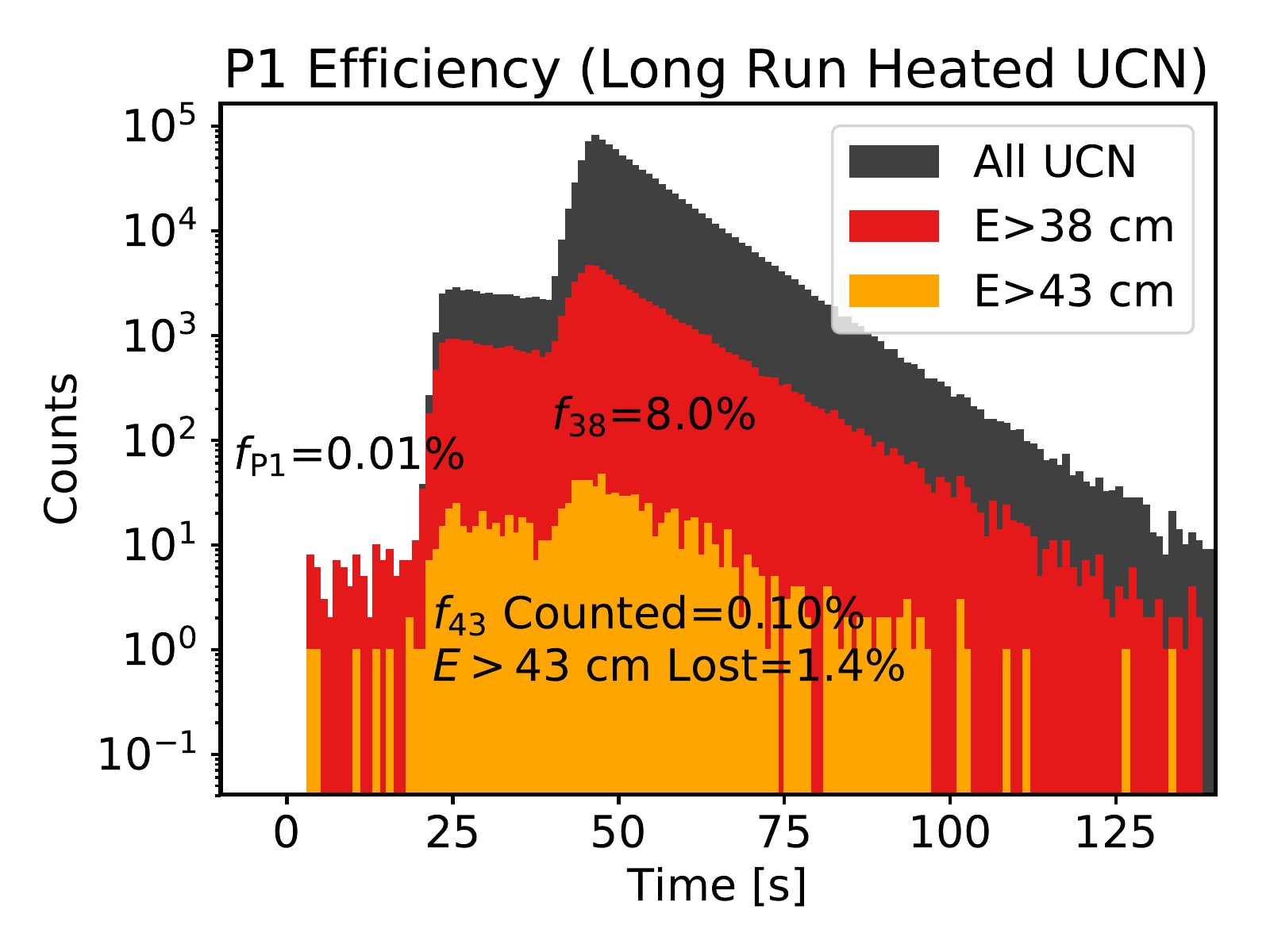}
    }
	\caption{The detection of high-energy neutrons ($E/m_ng>  38$~cm) in the 3-step measurement. The red histogram shows $E/m_ng> 38$~cm, the orange $E/m_ng> 43$~cm, and the gray shows all neutron counts, with \(f_{38}\) and \(f_{43}\) showing the integrated number of counted UCN of the corresponding energies. \(f_\text{P1}\) is the counts in Peak 1 normalized to the total population. Neutrons with $E/m_ng> 43$~cm could reach the cleaner at the raised height (during the long storage time) and be lost (without detection). 
(a) plots the 3-step data in a short-storage run to illustrates the detection of overthreshold neutrons due to insufficient spectral cleaning. (b) plots the same 3-step data in a long-storage run to illustrate the detection of overthreshold neutrons due to heating (of 80~$\mu$m amplitude). 1.4\% of UCN are lost to the cleaner at the raised height due to heating. }
	\label{fig:Heated}
\end{figure*}

If we attempted to correct for the effects of the heated neutrons, based on the counts registered in P1, then we need to consider the geometrical acceptance of overthreshold neutrons.
To accentuate the effects of heating (and get more statistics of heated neutrons), we increase the maximum vibration amplitude to 40~$\mu$m and 80~$\mu$m.
The corresponding shifts in lifetime due to heated UCN lost on the raised cleaner are \(\delta\tau=0.15\)~s and \(\delta\tau=7.68\)~s respectively. 
The results of the 80~$\mu$m heating simulation in a three-step detection experiment are shown in Fig.~\ref{fig:p1effHeated}.
The Monte Carlo simulations shows that the detector in the P1 does count the heated neutrons, however, a large population of the heated neutrons also appear in the lower two steps.
Based on the P1 subtraction procedure, we would have estimated that approximately ten heated UCN exist in the trap for every one UCN counted in P1 at the cleaning height. 
The Monte Carlo shows that the ratio is 750 of heated neutron for every UCN counted in P1, in the scenario with a large amplitude of microphonic vibrations.
For smaller amplitudes, the ratio gets smaller. Below a critical amplitude of microphonic vibrations, the heated neutrons slowly gain energy. The heated neutrons can extend into the region between the cleaning height and the raised cleaner (nominally the horizontal region with 5~cm in height), but they do not rise up to the raised cleaner. In this scenario, no neutrons are lost, and thus all neutrons counted in the multi-step detection---including the counts registered in P1---should be included to calculate the neutron lifetime.
This suggests that Monte Carlo input could be an important supplement for the P1 subtraction technique in setting a 
quantitative estimate (bound) of the heating effects.
We can attempt to put an upper bound using the population counted in step 1, however, more careful work is needed to estimate the efficiency of the dagger detector counting these heated neutron in the P1 position.

In the extreme heating scenario with 40~$\mu$m of vibration amplitudes, the Monte-Carlo predicts twenty UCN detected at P1, and this will lead to a lifetime shift of 0.15~s, which is below the current statistical precision.
In the experiment, the neutron count measured in P1 is consistent with background.
However, due to the size of the background, the data collected so far is statistically consistent with Monte Carlo simulations using 80~$\mu$m amplitude vibrations which would cause a shift in the lifetime of over 7 s. This is primarily due to low acceptance for counting heated UCN at the cleaning height.  Though such a large vibration amplitude is inconsistent with the accelerometer measurements.
With nominal vibration amplitudes, the Monte-Carlo simulation shows no measurable effect due to the heating loss. Once again, even though the P1 subtraction procedure might be insufficient to account for heating, given the small microphonic amplitudes, it does not have direct consequence to the neutron lifetime at the current level of precision.
Moving towards improving the precision to 0.1~s, 
it is necessary to implement long-term monitoring of the vibrations and
to apply necessary corrections using the Monte Carlo extracted geometrical acceptance to analyze the multi-step detection data.

\section{Summary}
We have developed an in-depth Monte Carlo model for the UCN\(\tau\) experiment. By using an analytical approximation of the magnetic fields, a thin-film detector model, and a parameterized spectral distribution of incoming neutrons, the simulations reproduce major features of the measured time distribution.
This model with five independent parameters, optimized using the data taken in the 9-step detection, qualitatively reproduces data taken in the 3-step detection. The model can be improved if the initial neutron distribution is better known. This can be achieved by a separate simulation that models detailed neutron transports from the production source to the UCN$\tau$ apparatus, with the trap door in the open position. Furthermore, efforts to calibrate the detector, to yield the information on the position-dependent scattering and absorption probabilities, will be needed to improve the agreement between the simulation to the data of neutron detection.
Nevertheless, the Monte Carlo model, calibrated by the neutron timing spectrum, was then used to explore the unique dynamics of the field-trapped neutrons.

In the simulations, by analyzing the rate of divergence between close trajectories, we categorize neutrons into chaotic (with large LCEs) and regular (with small LCEs). 
Low-energy neutrons are overwhelmingly regular.
Chaotic neutrons emerge with energies above 25~neV; even high energy neutrons, the population is roughly evenly divided between regular and chaotic. 
High energy neutrons started with polar angles $\theta_0>45^{\circ}$ are most likely to remain in regular, quasi-periodic orbits.
These neutrons present challenges to spectral cleaning, as it takes a long time for them to intersect with and be removed by the cleaner placed close to the top of the trap.
The current scheme of using a large-area cleaner suppresses the overthreshold neutrons (that could be lost on the cleaner at the raised height at 43~cm) below $10^{-4}$ of the total population of trapped neutrons. This limits the neutron lifetime shift due to insufficient spectral cleaning to below 0.03~s. 
Smaller corrections can be achieved by cleaning longer and/or deeper. 

Using the simulation, we also examine the multi-step detection and its power to infer the lifetime correction due to insufficient cleaning and heating, based on the counts registered in P1. 
We discover that the geometrical acceptance of the overthreshold neutrons, in particular the ones in regular orbits, is highly position dependent.
When the detector is placed in P1, the acceptance is small for the overthreshold neutrons in regular orbits. The geometrical acceptance increases in subsequent peaks as the detector is lowered, cutting into the phase space volume of these neutrons.
We also simulate neutron heating due to microphonic vibration, giving rise to a time-dependent magnetic fields. 
With nominal vibration amplitude of 1~$\mu$m, the lifetime correction is 0.03~s.
If the vibration amplitudes are anomalously large, then we might be able to infer the size of heating by counting the population of neutrons detected in P1. 
However, the P1 correction procedure, which assumes a constant geometrical acceptance and an ergodic neutron distribution, does not capture the underlying neutron dynamics.

Using this simulation, we also study details of the phase space evolution. 
Preliminary results, when compared to experimental data, show that the overall population of neutrons explore the phase space at a rate somewhat faster than the Monte-Carlo model predicts.
This suggests that the trapped neutrons either undergo excessive heating or the quasi-periodic trajectories evolve. This could result from irregularity of the magnets, or additional source of heating beyond the level of microphonic heating estimated in our experiment. 
Further work is needed to replicate the phase space evolution of the trap.

In conclusion, we have demonstrated, through simulations using a validated Monte-Carlo model, that the systematic effects of insufficient spectral cleaning and microphonic heating are no larger than 0.03~s. The UCN$\tau$ apparatus, coupled to the intense UCN source at LANL, has the potential to realize a neutron lifetime measurement to 0.1~s precision. 

\section{Appendix} \label{sec:appendix}

\subsection{Equations of Motion}
In the UCN$\tau$ trap, each neutron follows Hamiltonian mechanics, with a well-defined magnetic and gravitational potential energy. 
The trajectory of each neutron 
is tracked by numerically integrating Hamilton's equations:
\begin{eqnarray}
\frac{dp_i}{dt} &=& -\frac{d\mathcal{H}}{dq_j}  \mbox{; \vspace{5pt}} \frac{dq_i}{dt} = \frac{d\mathcal{H}}{dp_i} \mbox{; \vspace{5pt}} i=1,2,3 \\
\mathcal{H} &=& V(\bm{q})+\frac{\bm{p}^2}{2m_n} \\
V(\bm{q}) &=& -\bm{\mu}\cdot\bm{B}(\bm{q})+m_n g q_i \delta_{i3} = -\mu||\bm{B}||+m_n g z, \nonumber
\end{eqnarray} 
where $m_n$ and $\mu$ are the mass and the absolute magnitude of the neutron magnetic moment, $\bm{B}$ is the magnetic field, and $g$ is the local gravitational acceleration.
Since the neutrons are moving slowly in a continuous magnetic field, we use the adiabatic approximation, i.e., the angle subtended between the spin of the neutron and the magnetic field is preserved. Since the neutrons are polarized longitudinally, 
the neutron's magnetic moment, $\mu$, always anti-aligns with $\bm{B}$.
If the neutron passes through a fast varying magnetic field (or a time-varying magnetic field), at a rate comparable to or faster than the Larmor precession period, then the adiabatic approximation will not work. 
In this case, the spin would need to be tracked by numerically integrating the Bloch equations.

\subsection{Halbach Array field expansion}
The Halbach array consists of rows of permanent magnets arranged in strips.
The direction of magnetization between adjacent strips is rotated by $90^{\circ}$  (see Fig.~\ref{fig:trajintrap}, inset B).
These magnetic strips are placed along the major axis of the bowl, which is a part of a toroid. 
Magnetic field in the UCN$\tau$ trap, thus, can be described using a toroidal coordinate.
On the curved inner surface of the bowl, the local coordinates, $\eta$-$\zeta$-$\xi$, are designated with $\hat{\zeta}$ normal to the bowl surface, $\hat{\xi}$ along the major axis of the torii, and $\hat{\eta}$ along to the minor axis.
The coordinate $\zeta$ measures the distance from the array and $\zeta=0$ corresponds to the surface of the array.
Moving along the direction of the individual rows (in the $\hat{\xi}$ direction), the field is constant; moving along the direction crossing the rows (in the $\hat{\eta}$ direction, i.e.,  the minor axis of the toroid), the field rotates with a periodicity of 4 rows. The resulting field attenuates exponentially away from the surface of the magnets (in the $\hat{\zeta}$ direction).
We have mapped the field on the curved surface, and it follows the expected behavior of a Halbach array.
To calculate the force in the neutron-tracking simulation, we can use either the tabulated data from the field mapping or an analytical function that describes the field. 
The Halbach field inside the magnetic trap can be expanded in  a Fourier series~\cite{walstrom}\cite{berman}:  
\begin{eqnarray}
    \label{eq:expansion}
    \bm{B}&=&\frac{4B_\text{rem}}{\pi\sqrt{2}}\sum_{n=1}^{\infty}\frac{(-1)^{n-1}}{4n-3}\\*
    &&\times(1-e^{-k_nd})e^{-k_n\zeta}(\text{sin}k_n\eta\hat{\eta}+\text{cos}k_n\eta\hat{\zeta}), \nonumber
\end{eqnarray}
where \(B_\text{rem}\) is the remnant strength of the permanent magnets, \(k_n=2\pi(4n-3)/L\), with \(L\) the period of the rotated magnetization
, and \(d\) is the thickness of the magnet ($d$ = 25.4~mm). 
In the UCN$\tau$ trap,
\(B_\text{rem}\) is 1.35~T and \(L\) is 51.114~mm from empirical measurements.
Note that $L$ is slightly larger than its nominal 50.8~mm, due to the combined results of magnet size variation and finite gaps between magnets. Eq. \ref{eq:expansion} is an approximation that violates Laplace's equation in local bowl coordinates. However, since the field decays quickly compared to the curvature of the array the violation is $<1\%$.

To eliminate the possibility of neutron depolarization from field zeroes, the Halbach array field is superimposed with a holding field. The holding field is generated by a set of ten electromagnets, placed outside the vacuum chamber that houses the Halbach array. 
The resulting holding field is along the major axis of the toroid, $\hat{\xi}$; it is perpendicular to the field generated by the Halbach array. 
The orthogonality between the two fields ensures a non-zero magnetic field within the trap, with no accidental field cancellations.
The strength of the holding field is modeled by an ideal toroid
\begin{equation}
    B_\xi=B_0\frac{r+R}{\rho}\hat{\xi},
    \label{eq:holdingB}
\end{equation}
where \(\rho=\sqrt{x^2+y^2}\). $B_0$ is about 100 gauss with an applied current of 300~A in our coils. The origin of the lab frame is set 1.5~m above the lowest point of the trap.
The local coordinates $\eta$-$\zeta$-$\xi$ are related to the lab coordinates $x$-$y$-$z$ by
\begin{eqnarray}
    \xi &=& (R+r)\times \text{atan}\left(\frac{y}{z}\right)\\
    \eta &=& r\times\text{atan}\left(\frac{x}{\sqrt{y^2 + z^2} - R}\right) \\
    \zeta &=& r - \sqrt{x^2 + (\sqrt{y^2 + z^2} - R)^2}.
\end{eqnarray}
Here, \(R\) and \(r\) are the major and minor radii of the torus respectively. 
The total field strength is given by
\begin{equation}
\lVert\bm{B}\rVert=\sqrt{B_\eta^2+B_\zeta^2+B_\xi^2},
\end{equation}
with $B_\eta$ and $B_\zeta$ from the Halbach array, and $B_\xi$ from the holding-field coils.
The two halves of the magnetic bowl have \(r+R=1.5\)~m, but swap \(r\) and \(R\) at \(x=0\).
This feature breaks the left-right symmetry, to increase the number of chaotic neutron trajectories in the trap (see Sec.~\ref{sec:chaos}).
To smooth out the sudden change of radii along the middle line, we re-define the major and minor radii: 
\begin{eqnarray}
    R &=& \frac{1}{2} + \frac{1}{2\left(1 + \text{exp}(-\kappa x)\right)}     \label{eq:logr} \\
    r &=& 1 - \frac{1}{2\left(1 + \text{exp}(-\kappa x)\right)},
    \label{eq:logR}
\end{eqnarray}
where \(\kappa=1000\) is a parameter that sets the size of the crossover region near the \(x=0\) plane to approximately 5~mm.
\(R\) and \(r\) are chosen to be continuous functions of \(x\), in comparison to the values given by Ref.~\cite{walstrom}. This modification improves the degree of energy conservation in the simulation.

The total potential experienced by the neutron is proportional to \(\lVert\bm{B}\rVert\) and gravity. The minimum potential point (magnetic and gravitational) inside the trap was found at 3.5~cm above the trap bottom. This is the reference point for zero energy.
The forces on the neutron can be computed by taking the derivatives of the field:
\begin{equation}
    \frac{\partial\lVert\bm{B}\rVert}{\partial q_i}=
    \frac{1}{\lVert\bm{B}\rVert} \times \sum_{n=i,j,k}\sum_{m=i,j,k} B_{Q_n}  \frac{\partial B_{Q_n}}{\partial Q_m}\frac{\partial Q_m}{\partial q_i},
\end{equation}
where the lab coordinate $q_i = x, y, z$ for i=1,2,3, and the local coordinate  $Q_j = \xi, \eta, \zeta$ for j=1,2,3.
Here \(B_\zeta\) and \(B_\eta\) follow the expansion in Eq.~\ref{eq:expansion}.
To compute the component of the field in the lab frame, we take the derivatives of \(\eta\) and \(\zeta\) with respect to \(x,y,z\). These derivatives include the contributions from the logistic functions of Eqns.~\ref{eq:logR} and ~\ref{eq:logr}. The holding field \(B_\xi\) and its derivatives are calculated from Eq.~\ref{eq:holdingB}.

\begin{figure*}
    \subfloat[Energy variation of a UCN track over 10~s.\label{fig:10s500usencon}]{
        \includegraphics[width=0.45\textwidth]{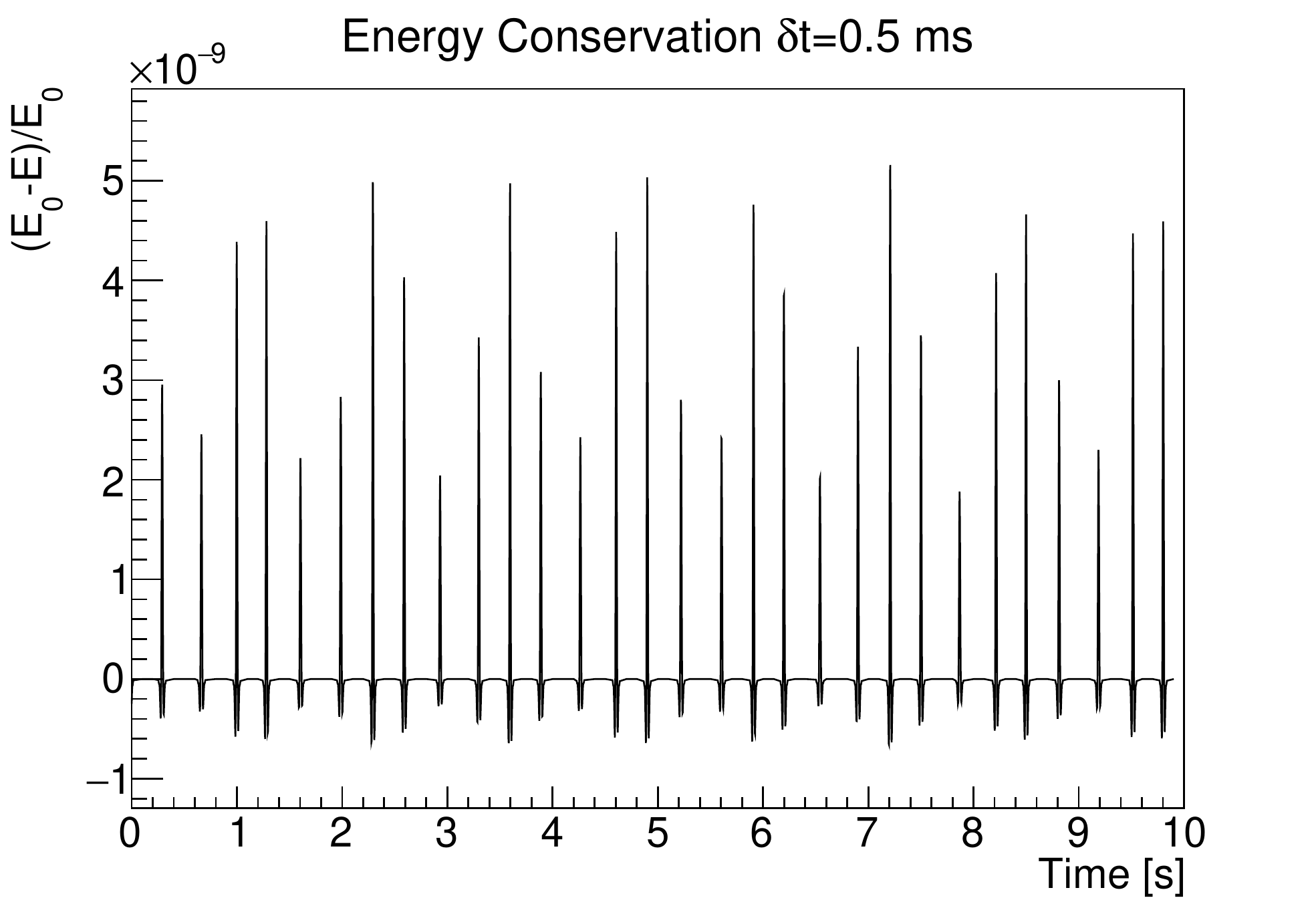}
    }
    \subfloat[Energy variation of a UCN track over 1000~s.\label{fig:1000s500usencon}]{
        \includegraphics[width=0.45\textwidth]{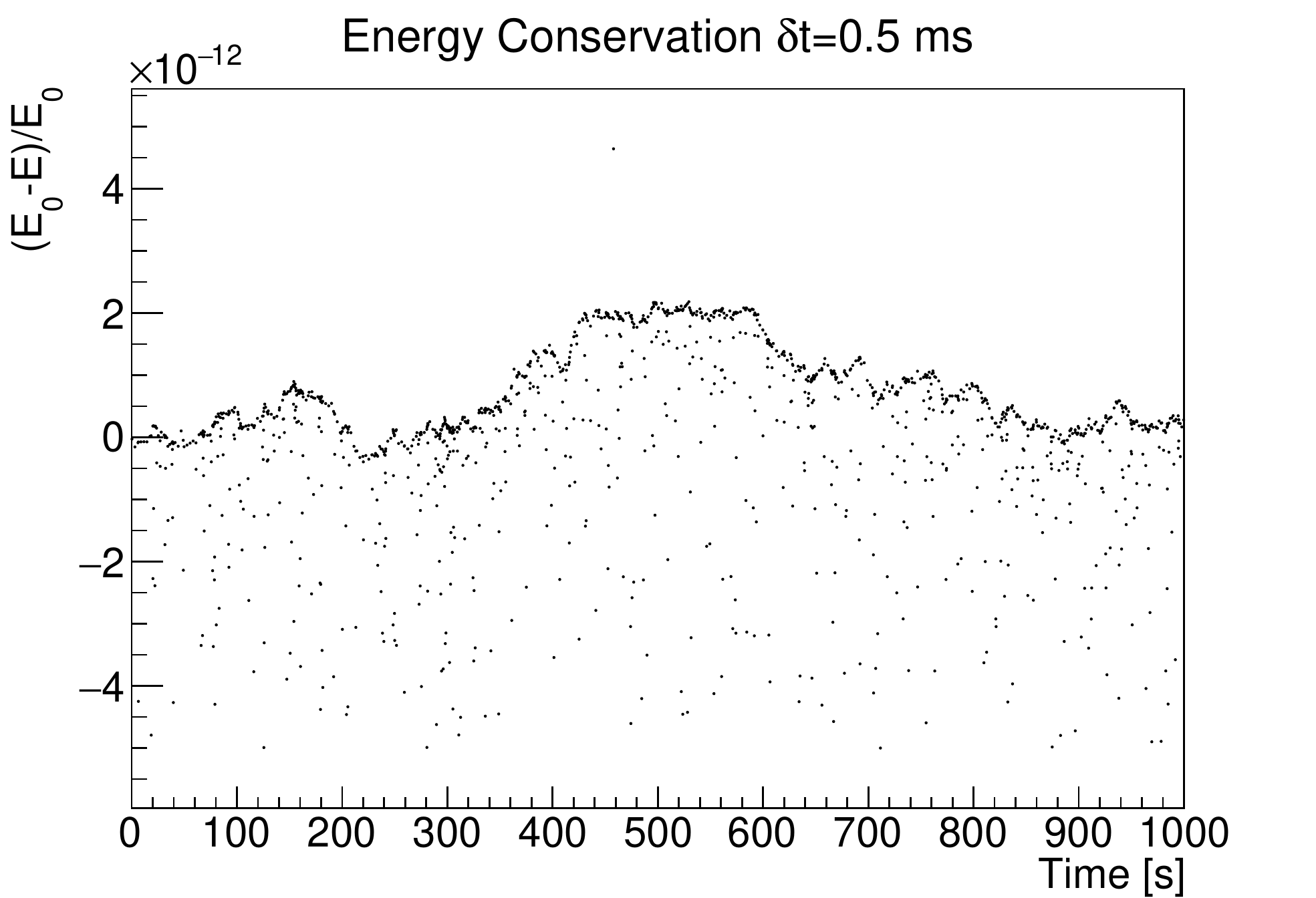}
    }\\
    \subfloat[Distribution of energy variation, \((E_0-E)/E_0\). The integration timestep is 0.5~ms.\label{fig:1000sdeltaE}]{
        \includegraphics[width=0.45\textwidth]{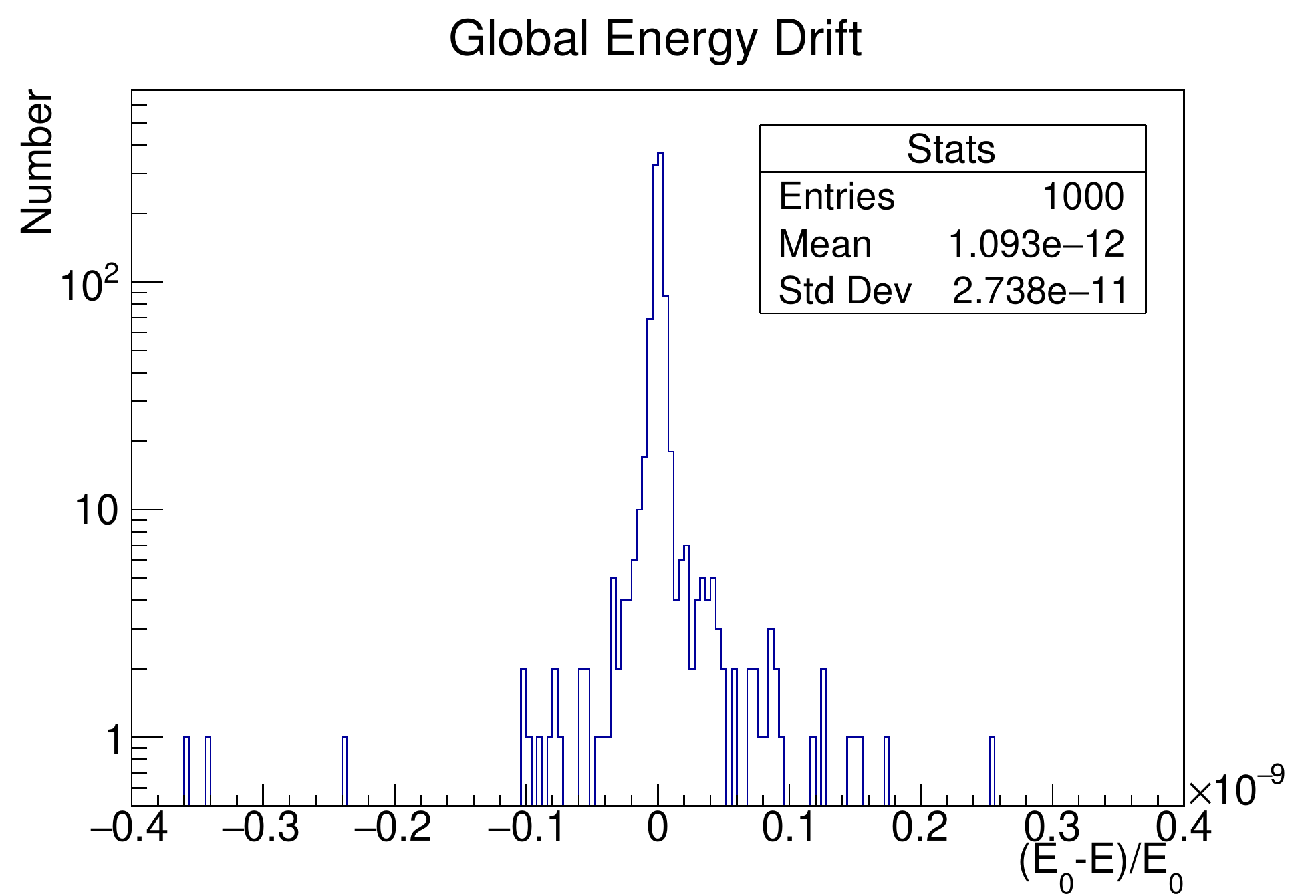}
    }
    \subfloat[Standard deviation of the energy variation  with varying timesteps.\label{fig:deltaEByStep}]{
        \includegraphics[width=0.45\textwidth]{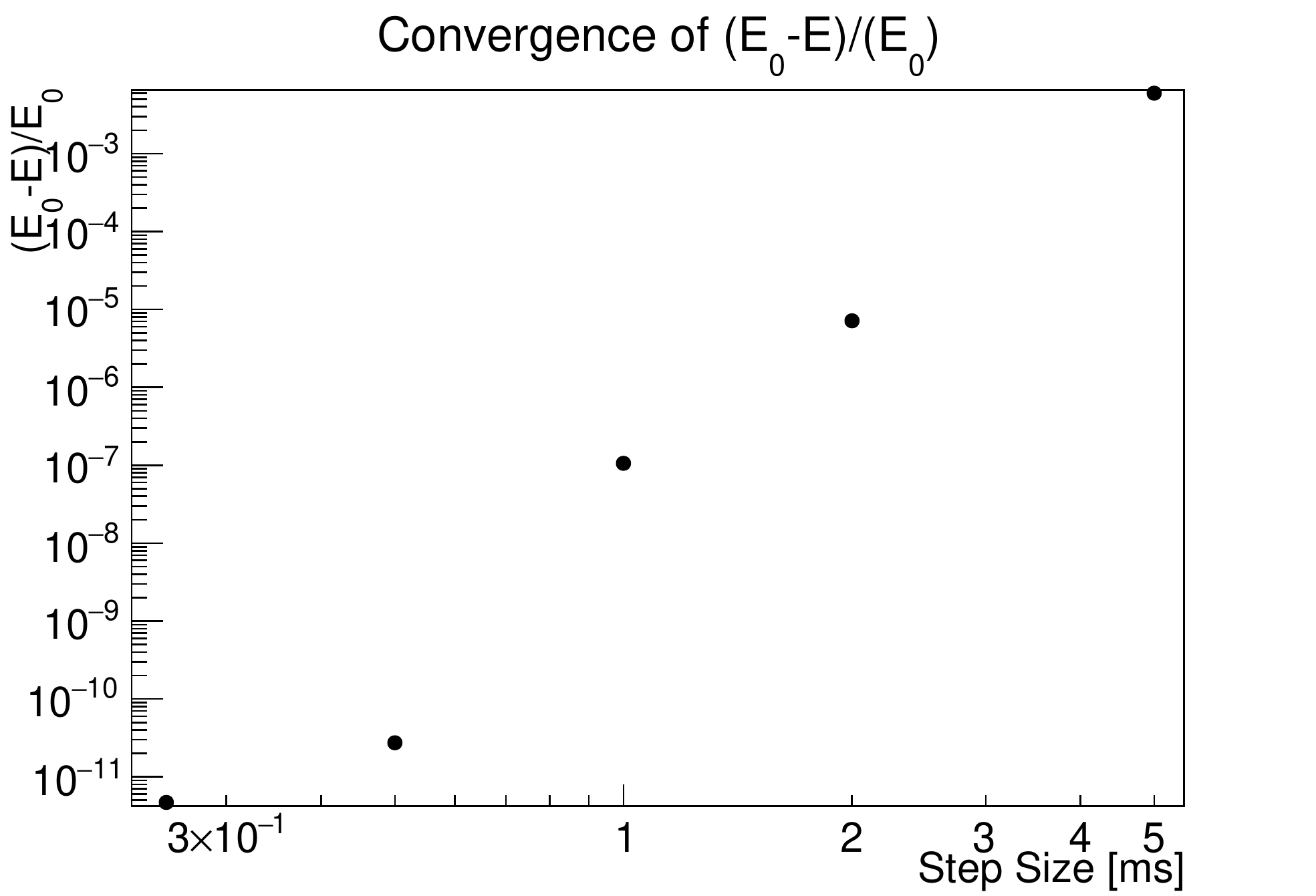}
    }
\caption{Study of the energy conservation of UCN tracks using the symplectic integrator. For subplot (c) and (d), the distribution is generated with 1000 UCN tracks, integrated over 1000~s. }
\label{}
\end{figure*}

\subsection{Symplectic Integration}
To solve the dynamics of this Hamiltonian system, a 4$^{th}$-order symplectic integrator is used to track individual neutrons inside the trap. 
The algorithm is given explicitly in Tab.~IV of \cite{candy}. This prescription gives an \(n\)$^{th}$-order symplectic integrator with time-dependent potentials, assuming the Hamiltonian is of the form \(H(\bm{q},\bm{p},t)=V(\bm{q},t)+T(\bm{p})\).
We use the coefficients given in Ref.~\cite{mclachlan},
which are numerically optimized to ensure better energy conservation than analytic coefficients.
With the force and the integrator, we evaluate the degree of energy conservation of individual neutrons by calculating the total energy for each step.
Fig.~\ref{fig:10s500usencon} illustrates the energy evolution of a sample trace over 10~s, numerically integrated with fixed timestep of 0.5~ms. 
The energy deviates when a neutron enters regions where the magnetic fields are highly non-uniform; a slight offset in its position can lead to a large difference in the potential energy. When the neutron moves back into low-field regions, the energy restores to its initial value. 
These spikes of energy deviations are the artifact of the numerical integration and should not affect the long-term stability of neutron tracking---a feature of the symplectic integration~\cite{hairer}.
The history of the global energy deviation of a typical simulated neutron trajectory is shown in Fig.~\ref{fig:1000s500usencon}. A histogram of \((E_\text{start}-E_\text{end})/E_\text{start}\) is given in Fig.~\ref{fig:1000sdeltaE} for a step size of 0.5~ms. The standard deviation of energy conservation as a function of step size is shown in Fig.~\ref{fig:deltaEByStep}.
The majority of the simulations presented in this work use a step-size of 0.5~ms; it gives a local deviation from the conserved energy on the order of \(10^{-8}\) and global drift on the order of \(10^{-11}\).

\begin{figure}
	\includegraphics[width=0.45\textwidth]{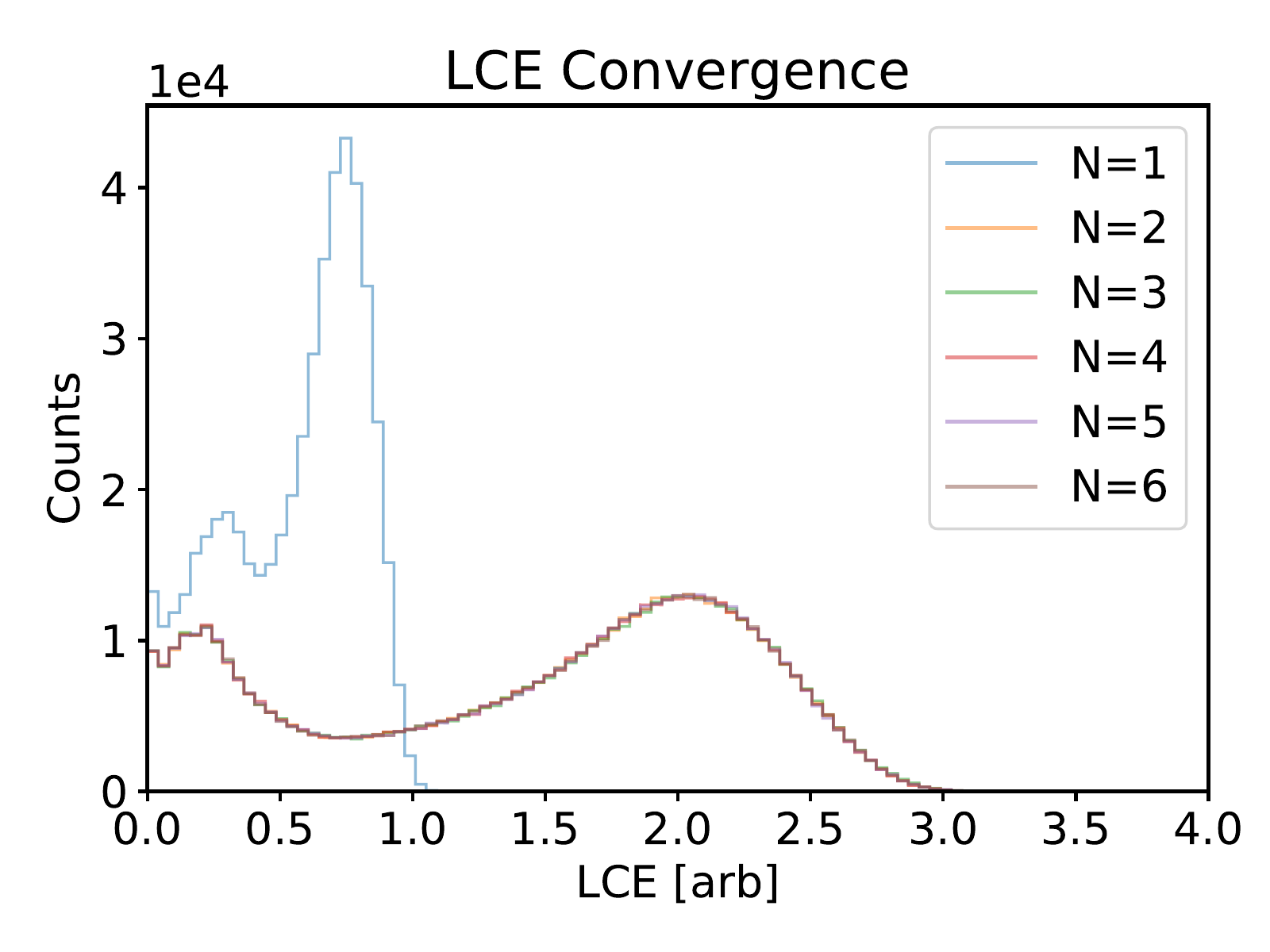}
	\caption{Distribution of LCE of simulated neutron trajectories using the field expansion Eq.~\ref{eq:expansion} up to $N$ terms.}
	\label{fig:LCEdistribution}
\end{figure}

To quantify the convergence of the results with step size and expansion cutoff,
we analyze the Lyapunov Characteristic Exponents (LCEs, defined in Sec \ref{sec:chaos}) of the simulated trajectories.
Individual simulated trajectories in other magnetic neutron traps do not converge with step size~\cite{coakley2005}, so we instead focus on the statistical distribution of the whole ensemble.
Adding more terms in the field expansion changes the Hamiltonian perturbatively.
Similarly, changing the timestep changes the exact Hamiltonian that is solved, and changes the evolution of a given initial condition.
However, if the distribution does not change with finer timesteps or more terms in the expansion, then we argue that the solution of the trap dynamics converges and the statistical behavior is stable at the macroscopic level.
 As shown in Fig.~\ref{fig:LCEdistribution}, adding more terms beyond 2 does not significantly change the distribution.
Furthermore, the number of UCN with Lyapunov Exponent $>$ 0.75 (considered chaotic trajectories) converges for $n>2$, with the fluctuations consistent with Poisson statistics. 
We therefore truncate the field expansion after 3 terms for the results presented here.
The distribution of Lyapunov exponents $>$ 0.75 is shown in Fig.~\ref{fig:LCEdistribution} to be singly-peaked. This indicates that we have measured the largest Lyapunov exponent in the system. A Hamiltonian system with $n$-dimensions contains $(n-2)/2$ independent Lyapunov exponents. The largest one will dominate the exponential growth, and therefore the measurement of the Lyapunov exponent using the method in Ref.~\cite{Benettin} should suffice to characterize the degree of chaos for each UCN trajectory.

\bibliography{bib}

\end{document}